\documentclass[iop,apj,tighten]{emulateapj}
\usepackage{apjfonts} 
\usepackage{amsmath,amstext}
\usepackage[breaklinks,colorlinks,citecolor=blue,linkcolor=magenta]{hyperref} 
\usepackage[all]{hypcap} 
\usepackage{threeparttable}
\usepackage{bm}
\usepackage{ulem}

\newcommand{\kms}{km$\rm s^{-1}$}
\newcommand{\masyr}{\,mas\,$\rm yr^{-1}$}
\newcommand{\mura}{$\mu_{\alpha*}$}
\newcommand{\mudec}{$\mu_{\delta}$}
\newcommand{\meanmura}{$\langle \mu_{\alpha*}\rangle$}
\newcommand{\meanmudec}{$\langle \mu_{\delta}\rangle$}
\newcommand{\mmura}{$\langle \epsilon_{\mu_{\alpha*}}\rangle$}
\newcommand{\mmudec}{$\langle \epsilon_{\mu_{\delta}}\rangle$}
\newcommand{\smura}{$\sigma_{\mu_{\alpha*}}$}
\newcommand{\smudec}{$\sigma_{\mu_{\delta}}$}
\newcommand{\dmura}{$\epsilon_{\mu_{\alpha*}}$}
\newcommand{\dmudec}{$\epsilon_{\mu_{\delta}}$}

\newcommand{\HJT}[1]{{{\color{black} #1}}}
\newcommand{\THJ}[1]{{{\color{black} #1}}}
\newcommand{\HJ}[1]{{{\color{red} #1}}}

\shorttitle{GPS1 proper motions}
\shortauthors{Tian, et al.}
\begin{document}
\title{A Gaia-PS1-SDSS (GPS1) Proper Motion Catalog Covering 3/4 of the Sky}
\author{
Hai-Jun Tian\altaffilmark{1,2}, 
Prashansa Gupta\altaffilmark{3},
Branimir Sesar\altaffilmark{2}, 
Hans-Walter Rix\altaffilmark{2},
Nicolas F. Martin\altaffilmark{2,5},
Chao Liu\altaffilmark{4},
Bertrand Goldman\altaffilmark{2,5},
Imants Platais \altaffilmark{6},
Rolf-Peter Kudritzki \altaffilmark{7},
Christopher Z. Waters \altaffilmark{7}
}

\altaffiltext{1}{China Three Gorges University, Yichang 443002, China.}
\altaffiltext{2}{Max Planck Institute for Astronomy, K\"onigstuhl 17, D-69117 Heidelberg, Germany. Email: hjtian@lamost.org}
\altaffiltext{3}{Department of Physical Sciences, Indian Institute of Science Education \& Research, Mohali , Sector 81, S.A.S Nagar, Punjab 140306, India.}
\altaffiltext{4}{Key Lab for Optical Astronomy, National Astronomical Observatories, Chinese Academy of Sciences, Beijing 100012, China.}
\altaffiltext{5}{Observatoire astronomique de Strasbourg, Universit\'e de Strasbourg, CNRS, UMR 7550, 11 rue de l'Universit\'e, F-67000 Strasbourg, France.}
\altaffiltext{6}{Department of Physics and Astronomy, The Johns Hopkins University, 3400 North Charles Street, Baltimore, MD 21218, USA.}
\altaffiltext{7}{Institute for Astronomy, University of Hawaii, 2680 Woodlawn Drive, Honolulu, HI 96822, USA.}

\begin{abstract}
We combine Gaia DR1, PS1, SDSS and 2MASS astrometry to measure proper motions for 350 million sources across three-fourths of the sky down to a magnitude of $m_r\sim20$\,.
Using positions of galaxies from PS1, we build a common reference frame for the
multi-epoch PS1, single-epoch SDSS and 2MASS data, and calibrate the data in small angular patches to this frame. As the Gaia DR1 excludes resolved galaxy images, 
we choose a different approach to calibrate its positions to this reference frame: we exploit the fact that the proper motions of stars in these patches are {\it linear}. By simultaneously fitting the positions of stars at different epochs  of --   Gaia DR1, PS1, SDSS, and 2MASS --  we construct an extensive catalog of proper motions dubbed GPS1. GPS1 has a characteristic systematic error of less than 0.3 \masyr\, and a typical precision of $ 1.5-2.0$\masyr. The proper motions have been validated using galaxies, open clusters, distant giant stars and QSOs. In comparison with other published faint proper motion catalogs, GPS1's systematic error  ($<0.3$ \masyr) should be nearly an order of magnitude better than that of PPMXL and UCAC4 ($>2.0$ \masyr). Similarly, its precision ($\sim 1.5$ \masyr) is a four-fold improvement relative to PPMXL and UCAC4 ($\sim 6.0$ \masyr). For QSOs, the precision of GPS1 is found to be worse ($\sim 2.0-3.0$\masyr), possibly due to their particular differential chromatic refraction (DCR). The GPS1 catalog will be released on-line and available via the VizieR Service and VO Service. (\HJ{GPS1 is available with VO TAP Query in Topcat now}, see http://www2.mpia-hd.mpg.de/$\sim$tian/GPS1 for details)

 \end{abstract}

\section{Introduction}\label{sect:intro}
Proper motions of stars in the Milky Way, along with precise distances and radial velocities, are important pieces of observational information. In particular, they are indispensable in building the six-dimensional phase space of these stars, which in turn provides vital information for understanding the kinematics of our Galaxy \citep{tian2015,tian2016,liu2016}. 

Several comprehensive proper motion catalogs have been released over the previous decade, which have improved the depth and accuracy each time. The PPMX catalog \citep{roser2008} includes proper motions with a typical precision of $2\sim10$\masyr for 18 million stars, down to a limiting magnitude of $\sim15$ in r-band. The PPMXL catalog \citep{roeser2010} uses a combination of the United States Naval Observatory B data \citep[USNO-B1.0;][]{Monet2003} and Two Micron All Sky Survey \citep[2MASS;][]{Skrutskie2006} astrometry. It includes objects to a magnitude of $V\sim20$, providing $\sim900$ million proper motions across the entire sky, calibrated to the International Celestial Reference System (ICRS); the typical individual proper motions uncertainties range from 4 \masyr\ to more than 10 \masyr , depending on observational history. \citet{vickers2016} made a global correction to the proper motions in PPMXL, taking care of the fact that extragalactic sources seem to originally have non-zero proper motions in PPMXL. \citet{zacharias2013} updated the UCAC series \citep{zacharias2004, zacharias2010} and published the latest release UCAC4. This catalog contains over 113 million objects covering the entire sky, of which 105 million have proper motions  complete down to about $R = 16$\, mag. 

Based on the Tycho-Gaia Astrometric Solution \citep[TGAS;][]{michalik2009}, the first data release of Gaia (Gaia DR1) published a catalog with proper motions in September 2016 for about 2 million Tycho-2 stars which only reach $G\sim12.5$\, \citep{hog2000,gaia2016a,gaia2016b}. Eventually, Gaia's proper motion measurements for more than a billion stars ($G\lesssim20.7$) in our Galaxy \citep{Bruijne2012,gaia2016a,gaia2016b} will
reach a level of 5-25 $\mu$as for $G\le 15$ stars, superseding all the previous ground-based measurements.  

While Gaia DR1 contained proper motions for  only 2 million TGAS stars, 
it also released precise J2015.0 positions for $\sim1$ billion stars across the entire sky \citep{gaia2016a,gaia2016b}. For 90\% of stars brighter than 19\, mag, the positional accuracies are better than 13.7\, mas, half of them are better than 1.5\, mas, and some even reach 0.1\, mas. 

Through more than five years of surveying, Pan-STARRS1 \citep[PS1;][]{Chambers2011,Magnier2016} has collected imaging data for billions of stars with high accuracy and multi-detections ($>60$ on average) for each source. The average uncertainty in positions is up to $\sim10$\, mas for stars brighter than 19\, mag in r-band. 

Here we set out to combine Gaia DR1's one-epoch position \HJT{measurement} at \HJT{very} high precision, with multi-epoch astrometry that PS1 survey provides, along with positions from SDSS and 2MASS at earlier epochs. This data set provides an unprecedented opportunity to build the best current proper motion catalog across much of the sky.

In general, two basic approaches can be used to bring proper motions to an inertial frame:
either \HJT{one can} use a highly accurate catalog that is already tied to the ICRS system, such as Hipparcos, and then add fainter sources to this system, as done for the Tycho-2 \citep{hog2000}, PPMXL \citep{roeser2010}, and UCAC4 \citep{zacharias2013} catalogs; or one can build a reference frame from distant extragalactic sources like galaxies (whose proper motions can be negligible), and cross-calibrate different epochs so that these sources have no proper motion. The proper motion catalog for SDSS \citep{munn2004} and the XPM catalog \citep{fedorov2009} were built using the latter method.

In this paper, we follow the second approach, and combine data from PS1, Gaia DR1, SDSS and 2MASS to obtain a catalog of proper motions dubbed GPS1. GPS1 is currently unmatched in its combination of depth, precision and accuracy among catalogs that cover a major portion of the sky. In Section 2, we detail the data sets involved. In Section 3, we lay out the approach for deriving reliable proper motions of stars from these surveys. We present our results, illustrating different data combinations, in Section 4, where we also validate the precision and accuracy of these proper motions with open clusters and distant halo stars, and make comparisons with published catalogs. We discuss possible problems that may induce small biases in proper motion estimates in Section 5. We conclude in Section 6.

Throughout the paper, we adopt the Solar motion as $(U_\odot,V_\odot,W_\odot)=(9.58, 10.52, 7.01)$\,\kms\ \citep{tian2015}, and the IAU circular speed of the local standard of rest (LSR) as $v_0=220$\,\kms. Also, $\alpha*$ is used to denote the right ascension in the gnomonic projection coordinate system, for example, \mura\ $=$ $\mu_{\alpha}\cos(\delta)$, and $\Delta\alpha*=\Delta\alpha\cos(\delta)$, while ${\epsilon}$ denotes uncertainties, to avoid confusion with the symbol $\delta$ referring to a source's declination. We use $\Delta$ to denote the differences in quantities such as proper motion or position.


\section{Data Set}\label{sect:data}
In order to construct proper motions, we analyze and model {\it catalog positions } 
from four different imaging surveys, \HJT{as discussed below}.
Gaia DR1 is based on observations collected between July 25, 2014 and September 16, 2015.
PS1 observations were collected between 2010 and 2014.
The SDSS DR9 data used here were obtained in the years between 2000 and 2008.
The images from 2MASS were taken between 1997 and 2001. The characteristics of the four astrometric catalogs are summarized in Table \ref{tab:surveys}.

\begin{table*}
 \begin{threeparttable}[b]
\caption{Characteristics of the four astrometric catalogs that constitute GPS1}.\label{tab:surveys}
\centering
\begin{tabular}{c|c|c|c|c|c|c}
\hline
\hline
Survey&Sky Coverage&Limiting Magnitude&Saturating Magnitude&Positional Uncertainty&Epochs& Average Detections\\
\hline
\multicolumn{2}{c|}{}&\multicolumn{2}{c|}{mag}&mas&\multicolumn{2}{c}{}\\
\hline
Gaia DR1&4$\pi$&$G\sim$20.7&$G\sim$11.2&$\sim$ 7&2015.0&1\\
PS1 PV3&3$\pi$&$r_{P1}\sim$22.0\tnote{a}&$r_{P1}\sim$13.5&$\sim$15&2010-2014&65\tnote{d}\\
SDSS DR9&$\pi$&$r\sim$23.1\tnote{a}&$r\sim$14.1&$\sim$25&2000-2008&1\\
2MASS&4$\pi$&$K_s \sim 14.3$\tnote{b}&$K_s \sim 8$\tnote{c}&$\sim$100&1997-2001&1\\
\hline
\hline
\end{tabular}
 \begin{tablenotes}
 \item [a] The limiting magnitude for detection with $S/N=5$.
 \item [b] The limiting magnitude for detections with $S/N=10$.
 \item [c] The saturating magnitude for detections in 1.3\,s exposure time.
 \item [d] The catalog of PS1 PV3, on an average, includes 65 detections for each source in $\sim$ 5 seasons.
  \end{tablenotes}
 \end{threeparttable}
\end{table*}

\subsection{Gaia}\label{sect:gaia}

After the first 14 months of observation, the ESA mission Gaia published its first data release (Gaia DR1) on September 14, 2016 \citep{gaia2016a,gaia2016b}. It consists of around 1.14 billion astrometric sources, of which only 2 million of the brightest stars contain the parallaxes and proper motions in the TGAS catalog (the so-called primary astrometric data set), while the other 1.1 billion sources have no proper motions (the so-called secondary astrometric data set). All the sources have positions and mean G-band magnitudes.

\begin{figure}[!t]
\centering
\includegraphics[scale=0.65]{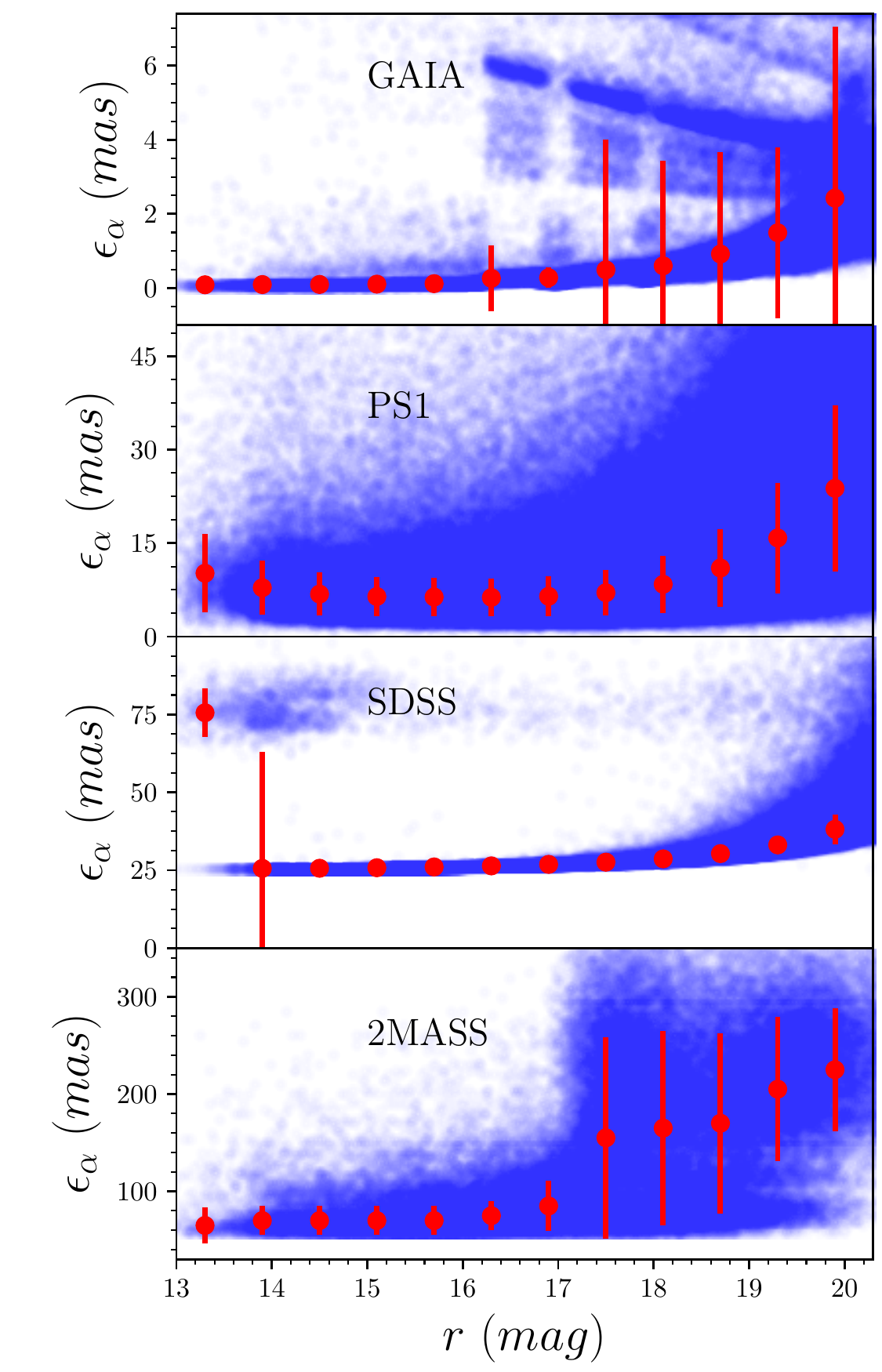}
\caption{Precision of the source position measurements for the various data used in the construction of the GPS1 catalog, as a function of $r$-band magnitude. The red dots and bars indicate the average and root-mean-square ({\it rms}) of the position uncertainties in each magnitude bin. Note the dramatically different vertical axis ranges of the different panels: the formal precision of Gaia DR1 (0.31 mas/year at $14<r<18$) is almost two orders of magnitude higher than 2MASS's. The PS1 precisions are about twice as good as SDSS.}\label{fig:mr_raErr}
\end{figure}

All the positions and proper motions in Gaia DR1 are calibrated to the International Celestial Reference Frame (ICRF) at epoch J2015.0. The typical uncertainty in positions and parallaxes, in the primary astrometric data set, is around 0.3 mas, while the TGAS proper motion uncertainties are around 1.0 \masyr. However, the proper motions for the $\sim$ 94000 Hipparcos stars are measured as accurate as 0.06 \masyr. The typical uncertainty in positions in Gaia DR1's secondary astrometric data set is $\sim$7 mas, as shown in the top panel of Figure \ref{fig:mr_raErr}. Note that $\sim$99.7\% sources in Gaia DR1 are in the magnitude range of $11.2<G<21$, as the saturating and limiting magnitudes are $G\sim12$ and 20.7, respectively \citep{gaia2016a}.

\subsection{Pan-STARRS1}\label{sect:PS1}
Pan-STARRS1 \citep[PS1;][]{Chambers2011} is a wide-field optical/near-IR survey telescope system, located at the Haleakala Observatory on the island of Maui in Hawaii. It has been conducting multi-epoch and multi-color observations over the entire sky visible from Hawaii (Dec $\gtrsim -30^\circ$). PS1 imaged in  five bands $g_{\rm P1}, r_{\rm P1}, i_{\rm P1}, z_{\rm P1}, y_{\rm P1}$, with a 5$\sigma$ single epoch depth of 22.0, 22.0, 21.9, 21.0 and 19.8 magnitude, respectively. The average wavelengths of its five filters are 481, 617, 752, 866, and 962 nm, respectively \citep{Stubbs2010, Tonry2012}. Unlike SDSS, PS1 observations in different bands are not taken simultaneously, and the wavelength coverage of the filters is also different. 
Roughly $56\%$ of the PS1 telescope observing time was dedicated to the PS1 3$\pi$ survey, which planned to observe each position 4 times per filter over 5 years. Throughout the 5 years, from 2010 to 2014, the PS1 3$\pi$ survey imaged a sky area of $\sim$30,000\ deg$^2$ in 65 epochs. Images are automatically processed using the survey pipeline \citep{Magnier2008,Magnier2016} that  performs bias subtraction, flat fielding, astrometry, photometry, as well as image stacking and differencing. The photometric calibration of the survey is $\sim 0.01$~mag \citep{Schlafly2012}.

All data processing shown here was carried under PS1 catalog Processing Version 3 \citep[PV3;][]{Chambers2016}. We stored the catalog locally in the Large Survey Database (LSD) format \citep{Juric2012}, which allows for a quickly and efficient manipulation of very large catalogs ($ > 10^9$ objects). The stored catalog contains both the point-spread function (PSF) and aperture photometry for each object, whose difference provides a convenient parameter for separating stars from background galaxies.

\subsubsection{Season Average and Positional Uncertainty in PS1}
The average number of total detections per PS1 source is 65, over 5.5 years. Each source is detected typically more than ten times in an observing season. We determine a robust average position and its uncertainties for each object within a season (hereafter, SeasonAVG). The typical single-epoch positional precision of bright ($r_{PS1}<19.0$) sources is $\sim 10\,$ mas, as illustrated in the second panel of Figure \ref{fig:mr_raErr}. 

\subsubsection{PS1 Astrometry Outlier Cleaning}\label{sect:outlier}
A comparison of position measurements among PS1 repeat observations for a sample of sources, shows that some estimates strongly deviate from the median position, and hence must be outliers.
To remove them, we apply selection cuts on (1) individual detections, and on (2) individual objects:
\begin{itemize}\itemsep1pt
    \item select detections for whom 85\% of their PSF lands on good CCD pixels ($psf\_qf>0.85$);
    \item remove detections with bad photometry ($photo\_flag\, \& \, 4027825560 = 0$), since problems that affect the PSF photometry also frequently affect the astrometry;
   \item remove detections that deviate by more than 3 times the robust {\it rms} scatter from their median values, where the robust {\it rms} is defined as $0.741\HJT{\times}(75^{th} \, {\rm percentile} - 25^{th} \, {\rm percentile})$ \citep{Lupton1993};
   \item keep objects with at least three 'good' detections;
   \item calculate the season-averaged position of PS1 astrometry
   (SeasonAVG), and its uncertainty from 'good' detections;
   \item keep objects with at least three SeasonAVG measurements. 
\end{itemize}

After the above filtering of the PS1 catalog, we keep 350 million objects with billions of detections. 

\subsection{SDSS}\label{sect:sdss}
The Sloan Digital Sky Survey (SDSS) used a dedicated 2.5-meter wide-field telescope \citep{Gunn2006} for imaging over roughly one third of the Celestial Sphere. The imaging was performed simultaneously in five optical filters: $u$, $g$, $r$, $i$ and $z$ with central wavelengths of about 370, 470, 620, 750 and 890\,nm, respectively \citep{gunn98, fuk96}. Stellar objects were uniformly reduced by the photometric pipeline. The $S/N=5$ limiting magnitudes are 22.1, 23.2, 23.1, 22.5 and 20.8 mag (AB system) in the five bandpasses, respectively. And stars saturate at 12.0, 14.1, 14.1, 13.8, and 12.3 mag in these same five bands. \citep{gunn98}

Since its regular operations began in 2000 April, SDSS has gone through a series of stages: SDSS-I \citep{York2000}, which was in operation through 2005, focused on a `Legacy' survey of five-band imaging and spectroscopy of well-defined samples of galaxies and QSOs, SDSS-II operated from 2005 to 2008, and finished the Legacy survey, followed by SDSS-III.  For our purposes, only the photometric data is relevant, especially the SDSS-I photometric sources which were imaged in the early epochs.  

The typical astrometric uncertainties for bright stars ($r<19.0$ mag) are around 20-30\,mas per coordinate \citep{Stoughton2002}, as shown in the third panel of Figure \ref{fig:mr_raErr}. 

While individual SDSS measurements are a factor of 2-3 less precise than PS1, the long epoch baseline makes this data very valuable.

\subsection{2MASS}\label{sect:2mass}
Two Micron All Sky Survey \citep[2MASS;][]{Skrutskie2006}, was conducted from two 1.3 m diameter dedicated telescopes located in the southern and northern hemisphere, which collected 25.4 terabytes of raw imaging data in the near-infrared $J$(1.25 $\mu m$), $H$(1.25 $\mu m$), and $K_S$(1.25 $\mu m$) bandpasses, covering virtually the entire celestial sphere between June 1997 and February 2001. The 2MASS All-Sky Data Release identifies around 471 million point sources, and 1.6 million extended sources. The limiting magnitudes at $S/N=10$ are $J=15.8$, $H=15.1$, and $K_s=14.3$, and point sources saturate at $K_s=8$ magnitude for less than 1.3\, s exposure time.

Bright source extractions have $1\sigma$ photometric uncertainty of less than $0.03$ mag and the astrometric accuracy is of the order 100\,mas, as shown in the bottom panel of Figure \ref{fig:mr_raErr}. Because of large positional uncertainties, 2MASS positions provide only a weak constraint for proper motion measurements.

\section{DERIVATION of PROPER MOTIONS}\label{sect:method}

The basic premise of our analysis is that the cataloged object coordinates, at any given
epoch, are precise relative coordinates of objects within a small angle on the sky (say, $\sim 1^\circ$). Yet, their absolute astrometry (i.e. the coordinates' {\it accuracy}) cannot be trusted across epochs and surveys. But all the ground-based imaging  surveys are deep enough to
contain a large number of compact or symmetrical galaxies with well-measured centroids,  for which the proper motions should
effectively be zero. We use those sources to bring the epochs to the same reference frame \citep[see, e.g.][]{munn2004}. While the Gaia imaging is of course deep enough to contain many galaxies, the positions of resolved objects have not yet been released in DR1. Therefore, we need a variant of the above
procedure to bring the ground-based data and Gaia DR1 to the same local reference frame.

\subsection{Qualitative Overview: Reference Frame Alignment and Proper Motion Fitting}\label{sect:methoddesc}
We give a brief summary of all the steps that lead to the construction of the proper motion catalog. For practical reasons, we consider different sub-areas of the sky in the course of this alignment:  a `tile' in this paper is an area of constant size of $10^\circ$ by $10^\circ$, a `patch' is a smaller region with area $1^\circ$ by $1^\circ$, and the 'pixel' is the smallest region with area $\sim$12 arcmin$^2$.

\begin{enumerate}
\item  \HJT{Select} a tile of \HJT{the} sky and acquire all \HJT{its} objects from PS1, Gaia, 2MASS, and SDSS (if it covers this region) databases.
\item  Classify the objects as stars and galaxies.
\item Separate the \HJT{tile} into equal-area pixels using the HEALPix system \citep{calrou2009} with 10 levels (i.e. $NSIDE = 10$), and label each pixel with its center position, namely, the Anchor Point (AP).
\item Construct a reference catalog by averaging repeatedly observed positions of PS1 galaxies in this \HJT{tile}. 
\item  Cross-match the PS1 objects with Gaia, 2MASS and SDSS using a $1.5\arcsec$ search radius.
\item  For each observing epoch, calculate the mean positional offset of galaxies relative to their reference position.
\item  Correct the positions of stars to the reference frame, assuming that their offset is the same as that of galaxies, in the same pixel and MJD.
\item To measure a proper motion of a star, fit a straight line (in the least squares sense) to PS1 SeasonAVG, 2MASS, and SDSS positions (if existing), where the positions are weighted \THJ{by} their inverse variance (the parallax is neglected).
\item \HJT{Use the information from $Step\, 8.$ to predict the stars' position at the Gaia epoch (2015.0). Then calculate the mean offset within a sky pixel between the stars' predicted positions and those of Gaia DR1.}
\item  \HJT{Use the offset from $Step\, 9.$ to bring the Gaia observations to the common reference frame.}
\item  Similar to $Step\, 8.$ fit the PS1 SeasonAVG, Gaia, 2MASS, and SDSS positions to get the final proper motion for each star.
\end{enumerate}
The following subsections detail the main steps in the above procedure.

\subsection{Reference and Astrometric Calibration}\label{sect:refercal}
We now elaborate more on the steps to bring the cataloged positions to a common 
reference frame before fitting for proper motions. 

\subsubsection{Sky-Direction Dependence of the Astrometric Offsets among Different Epochs}\label{sect:space_offset}

\begin{figure*}[!t]
\centering
\includegraphics[scale=0.67]{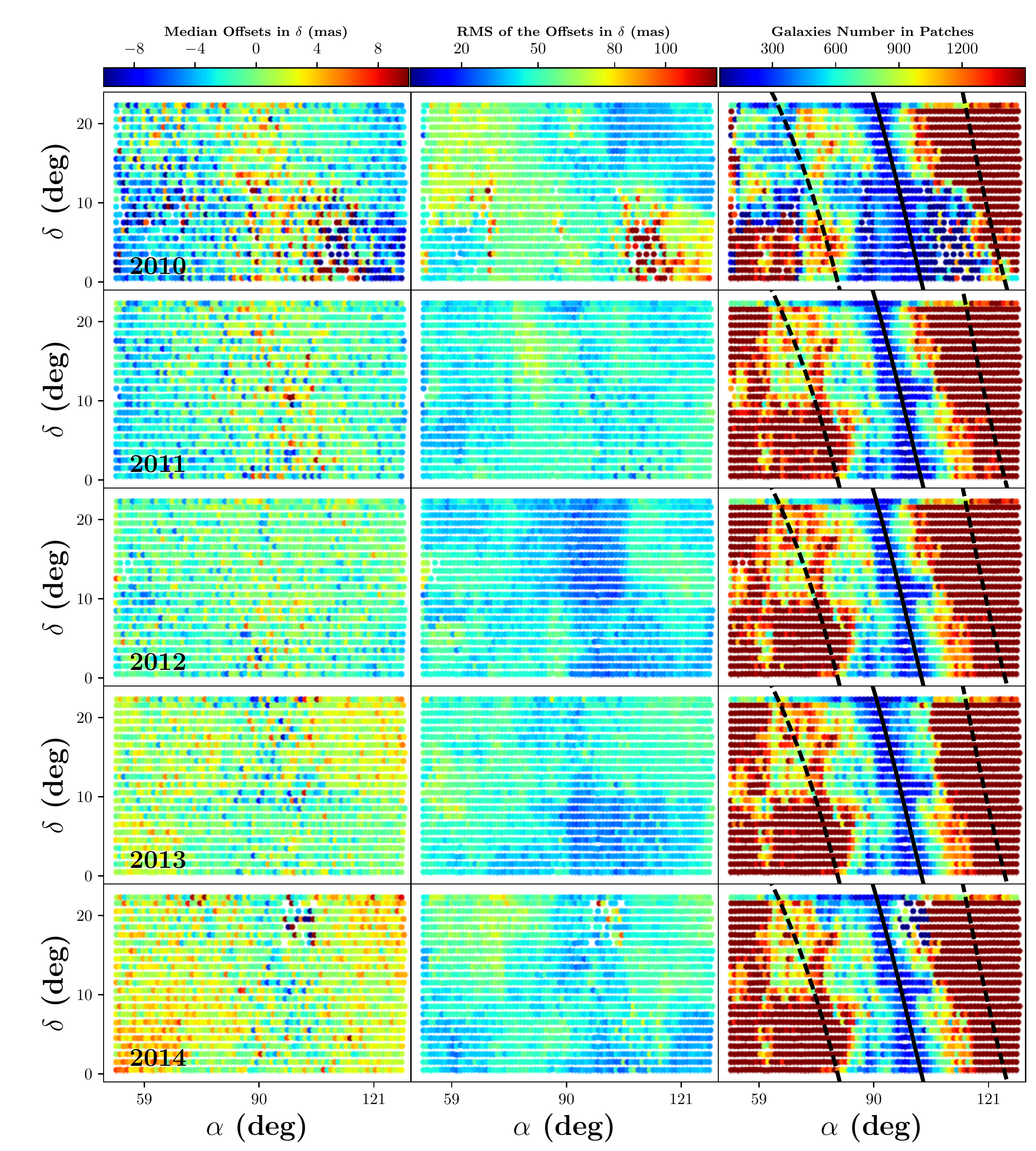}
\caption{Sample maps of the position offsets between individual PS1 epochs
and the mean reference positions, as a function of directions 
on the sky, shown for different epochs as different rows in the left column. 
The corresponding {\it rms} of positional offsets are shown in
the middle column, and the galaxy numbers used to determine the offset within each patch are shown in the right column. 
The median and {\it rms} of the offset for each patch are obtained from the positional residuals of all the galaxies within a patch relative to their median measurement of at least 3 epochs. Most of the patches include more than 600 galaxies, except for some regions close to the Galactic plane. The black lines in the right column mark the sky direction of $b=0^{\circ}$ (solid line) and $b=\pm 20^{\circ}$ (dashed lines).}\label{fig:offset}
\end{figure*}

We do not know {\it a priori} on what \HJT{angular} scales the positional offsets vary between different surveys and epochs. This must be determined from the data itself. Using PS1 data at relatively low Galactic latitudes,  we investigate the positional offsets of galaxies in different epochs, and find prominent offset patterns in different directions and epochs. Covering a sizable portion of the sky, Figure \ref{fig:offset} represents the median offsets among PS1 cataloged galaxy positions (the left column), along with the {\it rms} of the individual object's offsets (the middle column), and the galaxy numbers in each patch (the right column). The median and {\it rms} of the offsets are obtained from \HJT{all galaxies with at least three detections}. The black solid and two dashed lines correspond to the locations with Galactic latitude $b=0^{\circ}$, $-20^{\circ}$, and $20^{\circ}$, respectively. Different epochs of the same area in the sky are presented in the different panel rows. The offset and {\it rms} patterns remain unchanged if the patch size were changed to $0.5^{\circ}$ by $0.5^{\circ}$;
this leads us to choose $1^{\circ}$ as a radius to select background galaxies and use them to do the following calibration.

We then take \HJT{these median offsets in $\alpha$ and $\delta$, and add them} to positions of PS1 stars at a given epoch and in the current sky \HJT{pixel}. This is done separately for each sky \HJT{pixel} and different PS1 epochs. The single-epoch positions from 2MASS and SDSS are calibrated using the same procedure.

\begin{figure}[!t]
\centering
\includegraphics[scale=0.75]{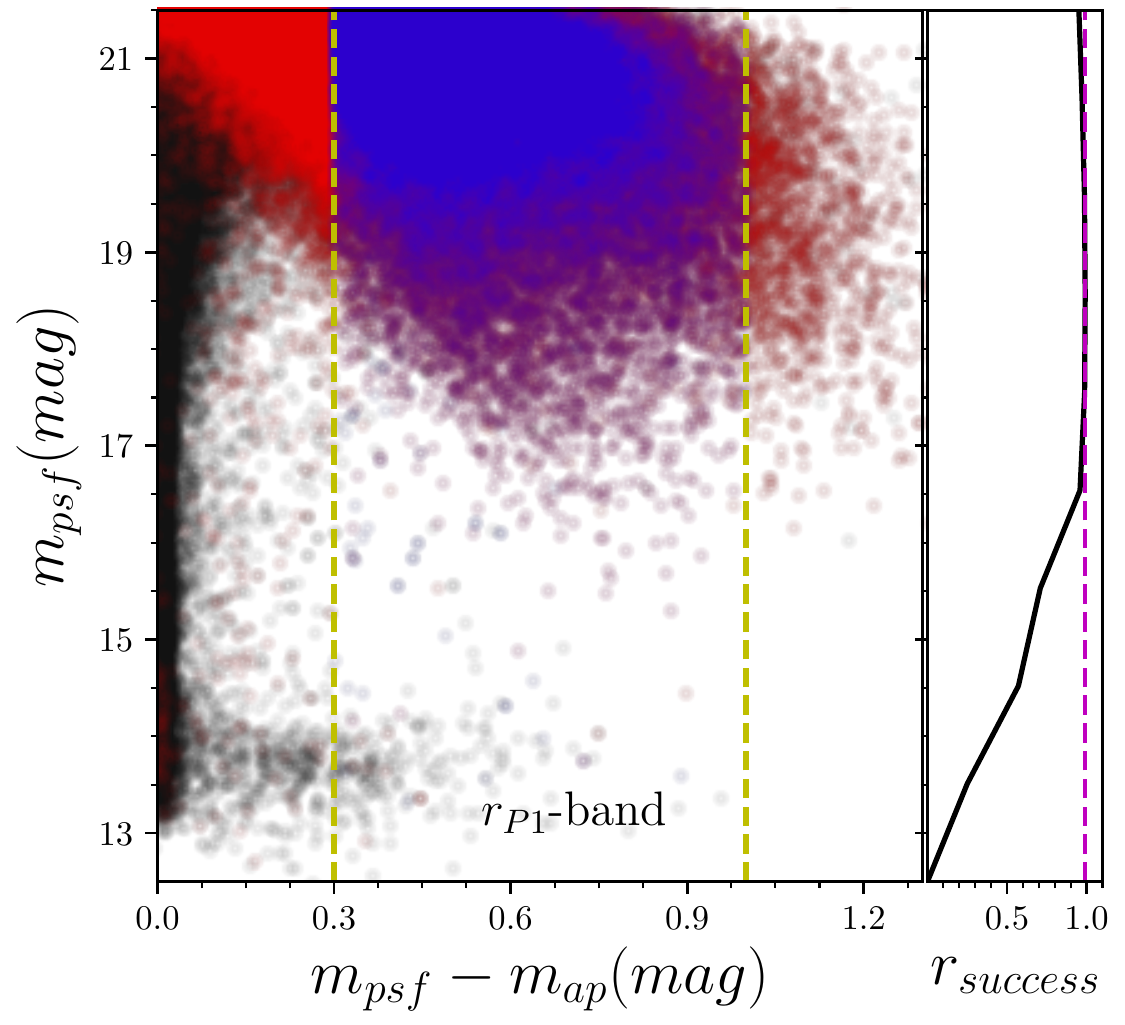}
\caption{The distribution of point and extended sources distribution in the plane of $m_{psf} - m_{ap}$ v.s. $m_{psf}$ in $r_{P1}$-band. Here, $m_{psf}$ and $m_{ap}$ are the PSF and aperture magnitudes, respectively. The figure illustrates star - galaxy separation \THJ{with PS1 photometry in the field of M67}, used to identify the galaxy sample for creating (and calibrating to) the position reference frame. The blue dots denote the galaxy candidates used for building the reference catalog, selected by $0.3\le m_{psf} - m_{ap}\le 1$ in both PS1 $r_{P1}$ and $i_{P1}$ (the borderlines are marked with the two yellow dashed lines). The red dots are galaxies identified by SDSS, which are assumed to be true galaxies. The black dots are the sources including point and extended sources. The right sub-panel displays the successful ratio ($r_{success}$) of this galaxy selection criterion. The ratio is obtained by cross-matching PS1 galaxy candidates (blue dots) and SDSS galaxies (red dots). This star-galaxy separation works very well for $m>16$, and  the ratio can reach up to 99\% (marked with the magenta dashed line).} \label{fig:gal_select}
\end{figure}

\HJT{The procedure described above requires careful identification of galaxies.} We define galaxies as objects for which the differences between point spread function (PSF) and aperture magnitudes in PS1 $r_{P1}$ and $i_{P1}$ bands lie between 0.3 and 1.0 mag. Using PS1 photometry in a field near M67, we can investigate how well this criterion works to selected galaxies. Sources \HJT{in the field of} M67 were observed and spectroscopically well classified by SDSS, and we take these classifications as the ground truth. Figure \ref{fig:gal_select} displays the distribution of point and extended sources in the panel of $m_{psf} - m_{ap}$ v.s. $m_{psf}$ in $r_{P1}$-band. The black dots are the sources from PS1 which \THJ{include} point and extended objects. The red points are the SDSS galaxies. The blue points are the galaxy candidates identified with the magnitude differences ($m_{psf} - m_{ap}$), which lies between 0.3\, mag and 1.0\, mag in both the $r_{P1}$ and $i_{P1}$ bands. By cross-matching with SDSS galaxies, we estimate the success fraction ($r_{success}$) of galaxy selection in different magnitude bins. The success fraction can reach up to 99\% for faint sources ($r_{psf}>17$ mag), as shown in the right sub-panel of Figure \ref{fig:gal_select}. This result indicates that galaxy selection criterion works fine for the selection of faint galaxies. In practice, the galaxies used to build the reference catalog in this work are dominated by faint galaxies. The galaxy candidates (blue dots) have higher contaminations at the bright end, mainly because of the image saturation. Even so, it \HJT{is still} safe to do the positional calibration using the median offset of hundreds of galaxies.

As Gaia's DR1 does not contain galaxies, we bring the Gaia positions to the common reference system, exploiting the fact that the proper motions of (almost) all stars are effectively linear. We use proper motions of bright stars ($14.5<m_r<17.5$) measured using PS1, 2MASS, and SDSS positions to predict the positions of the same stars at the epoch of Gaia observations (i.e., 2015.0). For the nearest 100 stars to each AP, we then take the median difference between Gaia's cataloged positions and the predicted reference frame positions at the given MJD. This offset is then subtracted from the Gaia positions of all stars located in that sky pixel.

We use simulations to validate this procedure for bringing the Gaia DR1 positions to our reference frame. We choose around 2000 stars from the PS1 catalog, and calculate their proper motions from PS1 detections. Using these proper motions, we predict the position of each star at Gaia's epoch, and record the positions as true locations of the simulated Gaia data. We divide the sky region into small equal-area patches. For each patch, we generate a random positional offset between -10 mas and 10 mas and assign the offset to each simulated Gaia star located in the same patch. For each star, we generate an additional random observational error ($\sigma=3.0\ $ mas). Finally, we calibrate the 'observed' Gaia stars with the nearby 20 stars, and calculate the differences between the true and calibrated positions. The median of the differences is around zero, and {\it rms} is smaller than 1.5 mas.

\subsubsection{Magnitude and Declination Dependent Offset Patterns}\label{sec:mag_offset}

\begin{figure*}[!t]
\centering
\includegraphics[width=0.4\textwidth, trim=0.0cm 0.0cm 0.0cm 0.0cm, clip]{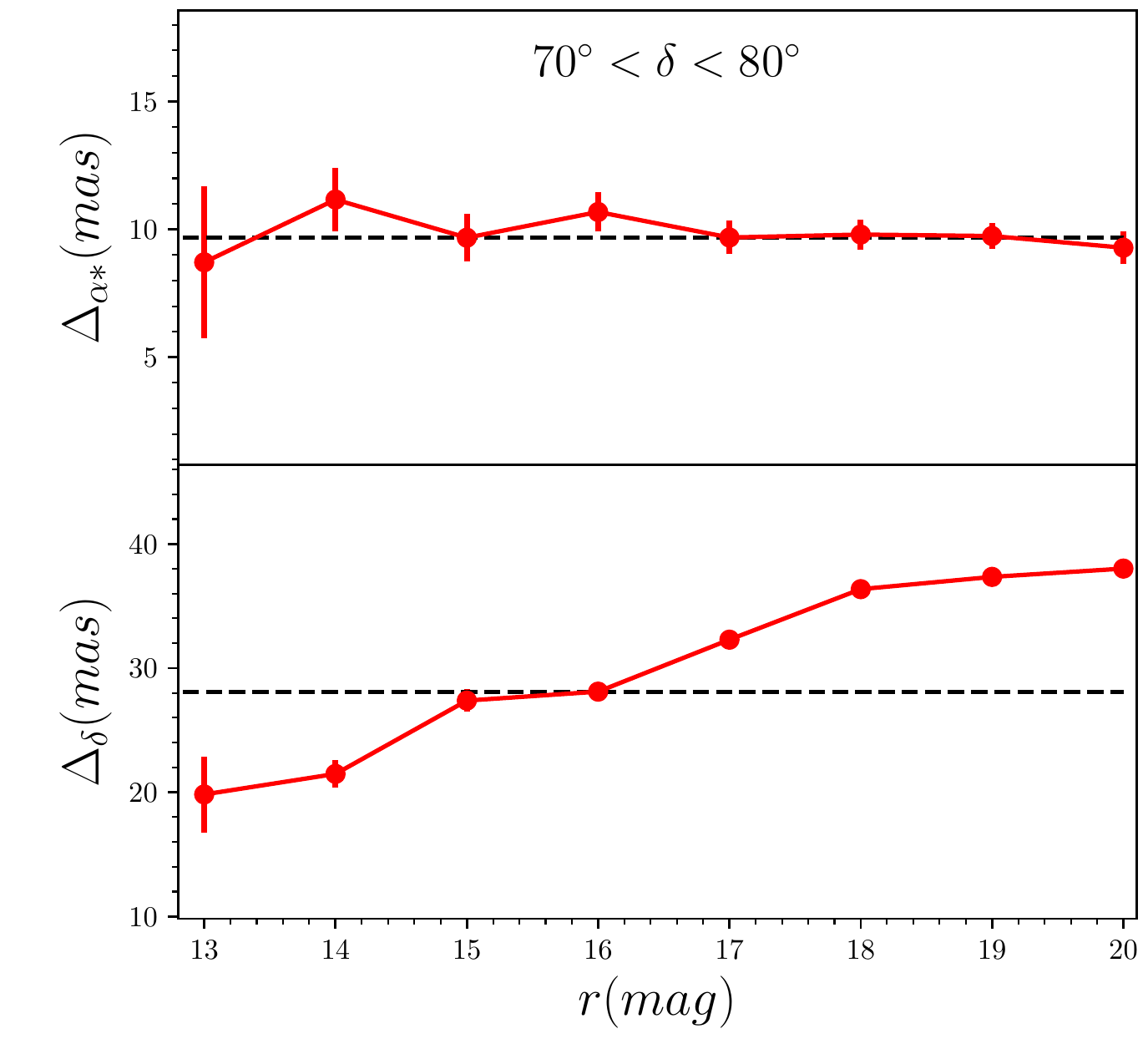}
\includegraphics[width=0.4\textwidth, trim=0.0cm 0.0cm 0.0cm 0.0cm, clip]{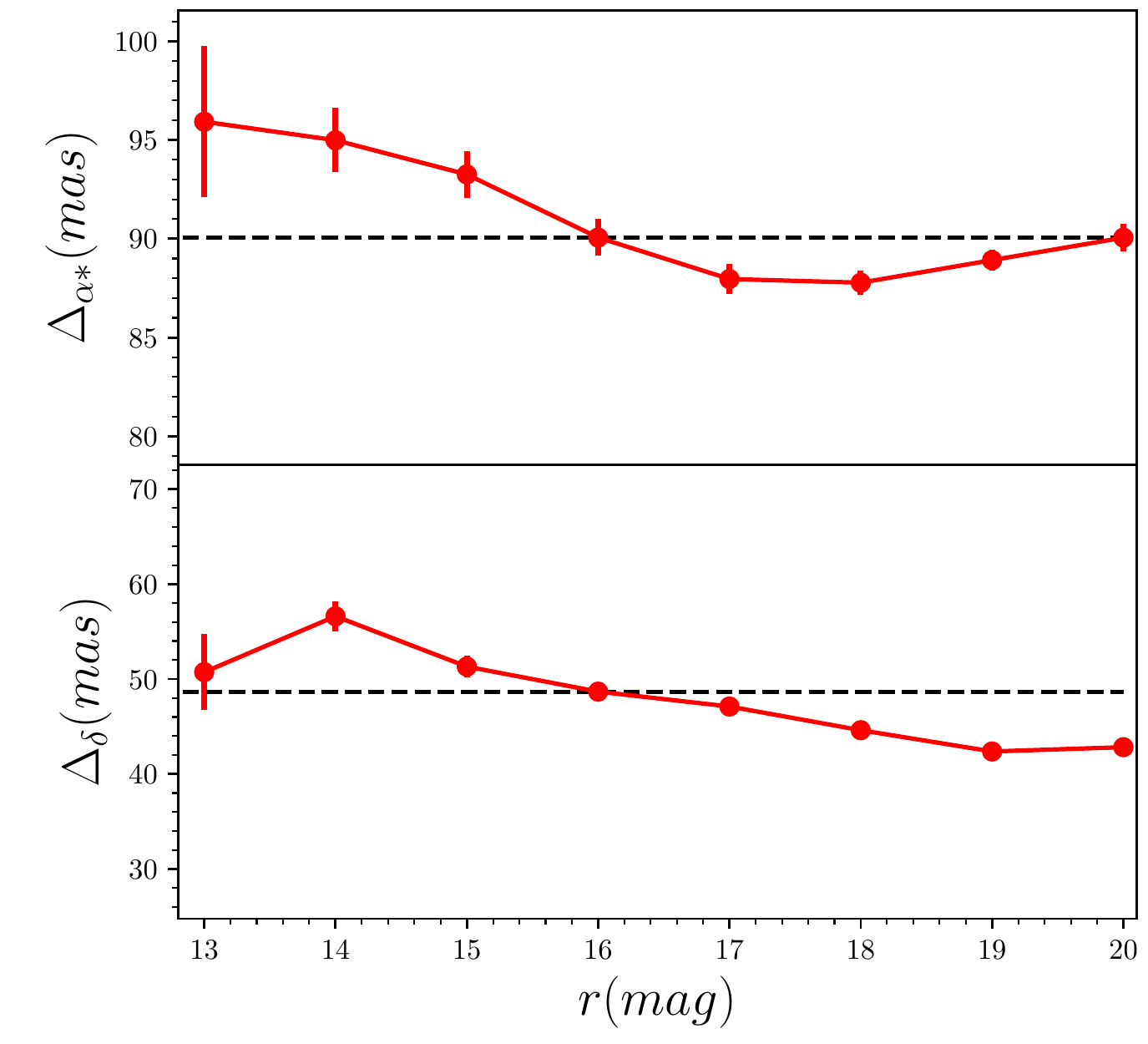}
\caption{The position offsets between Gaia and PS1 reference positions depend both on magnitude and on 
declination: high declinations  ($70^{\circ}<\delta<80^{\circ}$) are shown in the left panel and  low
latitudes in the right panel ($-30^{\circ}<\delta<-20^{\circ}$). The black dashed lines are the locations of median offsets with $14.5<m_r<17.5$\,mag, which mark the zero-point difference between Gaia and PS1-based reference. The red dots \HJT{and bars} are the median offsets and \HJT{uncertainties} in different magnitude bins. While the source of these trends presumably lies with the PS1 data, operationally we correct the Gaia positions to the PS1 reference positions, as we are only concerned with proper motions here.}\label{fig:offset2}
\end{figure*}

Even after \HJT{these} corrections, the differences between the PS1 reference positions and (corrected)
Gaia DR1 positions show some dependence on other quantities, namely, on declination and magnitude of the source. Figure \ref{fig:offset2} shows the variation of the mean positional offsets with r-band magnitude, both at high (the left panels) and low (the right panels) declinations. The positional offset for each star is obtained by taking the difference of the Gaia's predicted position and the originally observed position. The predicted position for each star is calculated from the PS1, SDSS, and 2MASS fitted proper motion. The black dashed lines are the locations of the median offsets with $14.5<m_r<17.5$, which mark the zero-point difference between Gaia and PS1-based reference. The red dots \HJT{and bars} are the median offsets and \HJT{uncertainties} in different magnitude bins, \HJT{and} they \HJT{show} obvious variations with magnitude. In particular the high and low declination variations in the direction of $\delta$ (the bottom panels) are almost opposite, while in the direction of $\alpha$ (the top panels) the offsets keep roughly constant. Irrespective of the origin of this offset pattern, it \HJT{must be} removed or mitigated. To do so, we build a relation model between the offset and magnitude on a larger angular scale, i.e., for each sky tile. For most tiles, the offset is roughly linear with magnitude,
\begin{equation}\label{eq:pdelta}
\Delta(\delta, m) =  c\cdot m - \Delta(\delta, m_0),
\end{equation}
where $\Delta(\delta, m_0)$ is the zero-point difference between the Gaia and PS1-based reference at a given declination (the dashed lines in Figure \ref{fig:offset2}), $m_0$ is the average magnitude of stars with $14.5<m_r<17.5$ (since we use the stars in this magnitude bin to place Gaia positions onto the PS1 reference frame), and $c$ is the slope of the offset line. In practice, one could also remove the magnitude dependent offset for each star by linear interpolation \HJT{between the magnitudes and offsets}.

While we have been able to correct for this effect, we have not been able to
identify its root cause with any certainty. It seems plausible that 
it can be traced to the PS1's experimental set-up:  it is known that the 
cataloged PS1 positions have had some  magnitude dependence \citep{Koppenhofer2011}, and the differential chromatic refraction (DCR) \citep{Kaczmarczik2009} may be imperfectly corrected. This issue will be discussed further in Section \ref{sect:moffset}. This type of offset is also detected in SDSS, but at a much lower level ($<5$ mas). The offset in SDSS positions does not significantly affect the final proper motion measurement as its is much smaller than the average positional uncertainty in SDSS positions ($\sim 25$ mas).

\subsection{Proper Motion Fitting}\label{sect:fitting}
After calibrating the cataloged positions for each object in five (or six) PS1 epochs, one Gaia epoch, one 2MASS epoch, and possibly one SDSS epoch onto the same reference frame, we can calculate the proper motion for each star by performing a linear least squares fit to positions observed at up to nine different epochs.
We do this by using a simple $\chi^2$ fit that includes outlier rejection. We start with 

\begin{equation}\label{eq:chi2}
\chi^2=\sum_{i}^{N}\frac{[\hat{y}_{i}^{o} - y_{i}^{model}(t_i)]^2}{\epsilon_{i}^2},
\end{equation}
where $\hat{y}_i^o$ is the observed position of a star at epoch $i$, and
$\epsilon_{i}$ the position uncertainty. All the positional uncertainties consist of two parts: one part is the individual position precision, illustrated in Figure \ref{fig:mr_raErr}; and the other part is the uncertainty from the offset calibration, illustrated for PS1 in Figure \ref{fig:offset}. $y_i^{model}(t_i)$ is the predicted position by a linear model at the given time $t_i$, $N$ is the number of epochs in different surveys.
The position $\hat{y}_i^o$ has been calibrated by
\begin{equation}\label{eq:cal_offset}
\hat{y}_i^o = y_i^o - \Delta_{i}(\alpha, \delta) - \Delta_{i}(\delta, m),
\end{equation}
where $y_i^o$ is the original cataloged position of a star at epoch $i$, $\Delta_{i}(\alpha, \delta)$ is the direction dependent offset described in Section \ref{sect:space_offset}, and $\Delta_{i}(\delta, m)$ is the magnitude and declination dependent offset described in Section \ref{sec:mag_offset}.  

\begin{figure*}[!t]
\centering
\includegraphics[width=0.45\textwidth, trim=0.0cm 0.0cm 0.0cm 0.0cm, clip]{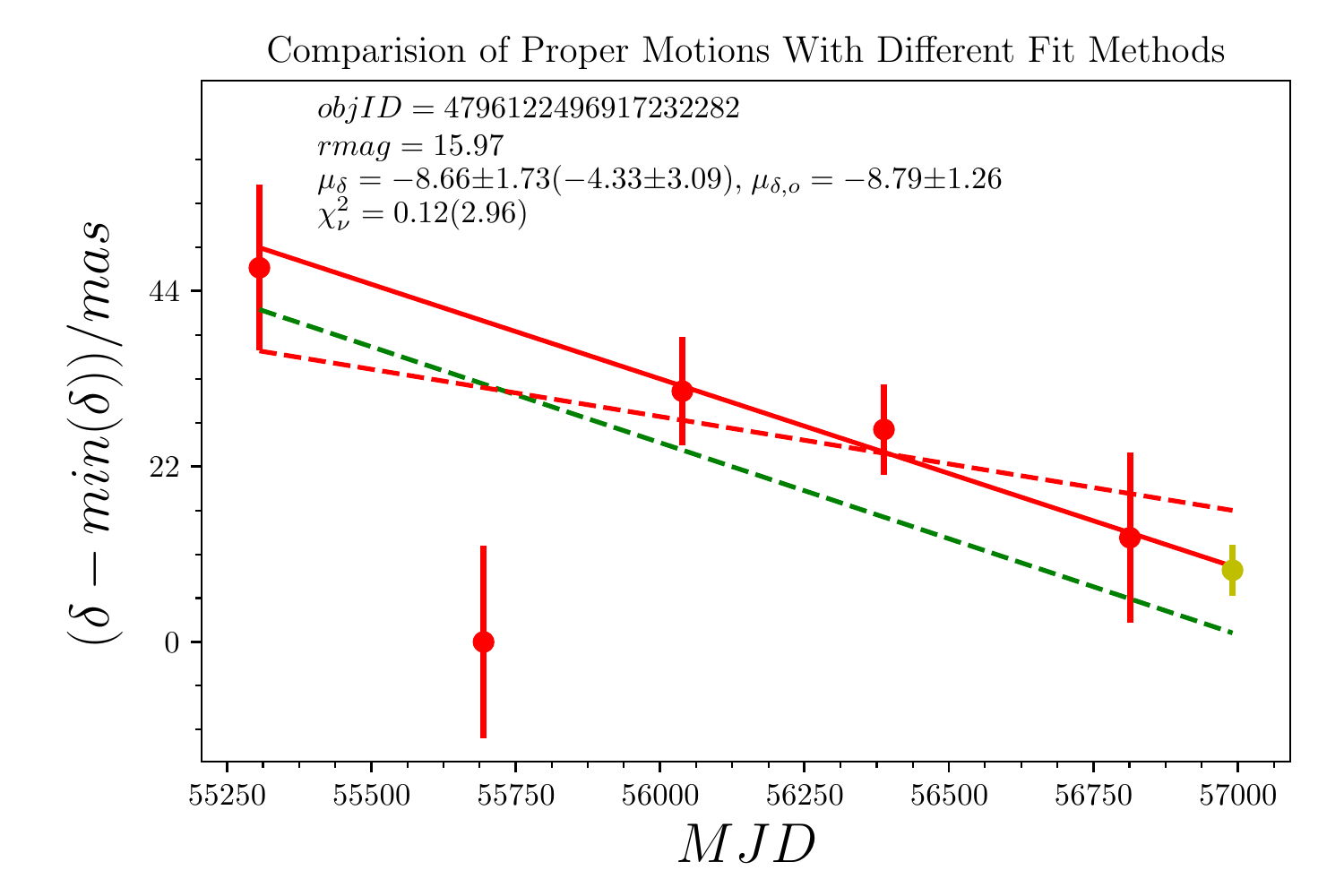}
\includegraphics[width=0.45\textwidth, trim=0.0cm 0.0cm 0.0cm 0.0cm, clip]{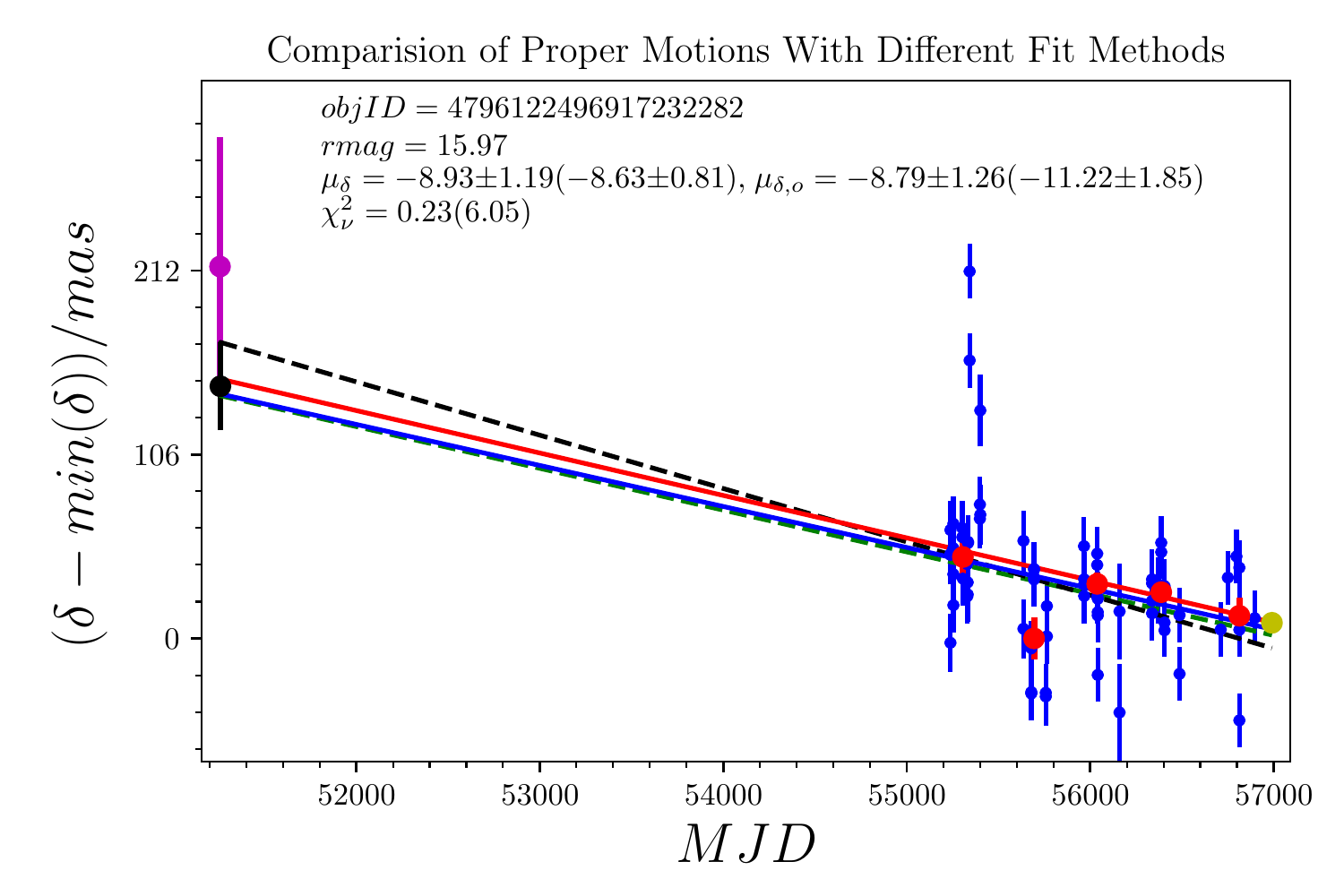}

\caption{Illustration of the different options in proper motion fitting, using the various combinations of data sets,
and outlier rejection. The left panel shows fits to the combination of the Gaia (yellow) and PS1 SeasonAVG data (red). The red point at MJD$\sim$56800 in the PS1 SeasonAVG data is an obvious outlier. The red solid line shows the proper motion estimate, $-8.66\pm1.73$\masyr, after removing one PS1-season outlier and including the Gaia DR1 (yellow) point. The red dashed line shows the analogous estimate, $-4.33\pm3.09$\masyr, when fitting all five PS1 epochs, including the outlier, but without the Gaia DR1 epoch. In the right panel, the red solid line (proper motion: $-8.93\pm1.19$\masyr, 4 PS1 SeasonAVG excluding the outlier) and blue solid line (proper motion: $-8.63\pm0.81$\masyr, 64 PS1 individual blue points) are drawn by fitting the combination of Gaia (yellow dot), PS1, SDSS (black dot), and 2MASS (magenta dot). The black dashed line (for comparison) on the right is plotted according to the proper motion ($-11.22\pm1.85$\masyr) from PS1 PV3 catalog. The green lines in both panels are according to the proper motion ($-8.79\pm1.26$\masyr) from by \citet[][]{fk2015}. The fitting is taken on a same star example in the left and right panels.
 }
   \label{fig:fitting}
\end{figure*}

Unrecognized outliers in positional data may induce a spurious proper motion estimate. In order to remove such outliers, we employed leave-one-out cross-validation. We withhold one of the observation epochs, fit a straight line to the remaining positions, and calculate the reduced $\chi_{\nu}^2$ ($\equiv \chi^2/(N_{data\ points}-2)$). This procedure is repeated for each observation epoch, and we adopt the fit with the minimum $\chi_{\nu}^2$. In practice, leave-one-out fits can eliminate outliers efficiently. The left subplot in Figure \ref{fig:fitting} represents a typical SeasonAVG outlier (the second red point), which severely affects the proper motion fitting, as \HJT{shown} by the red dashed line. 

Even though Gaia DR1 contributes only one epoch, precisely anchoring down the position at that one epoch can significantly reduce the proper motion uncertainties. For example, the red solid line in the left subplot of Figure \ref{fig:fitting} illustrates the proper motion ($-8.66\pm1.73$\masyr) by fitting the 4 PS1 SeasonAVG points (the red dots, excluding the outlier) and 1 Gaia point (the yellow dot). The uncertainty of the proper motion is reduced by $\sim1.36$\masyr, compared to the fit based on PS1-only, represented by the red dashed line.

The typical positional uncertainty of SDSS is around 25\, mas for objects brighter than 19\, mag, an order of magnitude worse than Gaia. Yet, the
long epoch-baseline of SDSS makes these data important for the proper motion fit. The black point in the right hand panel of Figure \ref{fig:fitting} is a position observed by SDSS about 15 years ago. The proper motion ($-8.93\pm1.19$\masyr) represented by the red solid line is obtained by fitting the 4 SeasonAVG points (the red dots, excluding the outlier), 1 Gaia point (the yellow dot), 1 SDSS point, and 1 2MASS point (the magenta dot), simultaneously. Compared to $-4.33\pm3.09$\masyr\ (only PS1), the uncertainty is reduced by $\sim1.9$\,\masyr; including the 2MASS point actually improves only by $\sim0.05$\, \masyr. Besides improving the individual objects' precision, the SDSS and Gaia points also enhance the accuracy of the proper motion. Wherever SDSS is available, the 2MASS point contributes little weight because of its large positional errors (on average $>100$\, mas for objects brighter than 20\, mag).

The blue points in the right-panel of Figure \ref{fig:fitting} show the 64 individual detections from PS1 (as opposed to the seasonAVG points). These points have been cleaned of outliers by the cuts described in Section \ref{sect:outlier}. The blue solid line (proper motion $=-8.63\pm0.81$\masyr) is obtained by fitting 67 points (64 individual PS1, 1 SDSS, 1 Gaia, and 1 2MASS) simultaneously, consistent within $1\sigma$ with the red line (by fitting 4 SeasonAVG, 1 SDSS, 1 Gaia, and 1 2MASS points). The positional uncertainty cannot be estimated straightforwardly for any one individual detection. Therefore, we used a simple empirical model to assign the positional uncertainty for each star according to its magnitude,
\begin{equation}\label{eq:error}
\sigma_{(\alpha, \delta)} (mas)=\sqrt{15^2 + (1000\times\epsilon_{m_r})^2},
\end{equation}
where $\epsilon_{m_r}$ is the $r$-band magnitude error in the individual detection. This formula can model the relation between observational uncertainty and magnitude, but cannot well discriminate among uncertainties for the same object in different detections. Therefore, the modeled observational uncertainty cannot qualify as weight in the proper motion fitting procedure. That means, the precision (0.81\,\masyr) of the proper motion obtained by fitting the blue points is \HJT{unreliable}.

For comparison, the proper motions from other catalogs for the star in the fitting example are displayed in Figure \ref{fig:fitting}. The green dashed line is the proper motion estimate ($-8.79\pm1.26$\masyr) from \citet{fk2015}. The proper motions fitted with either seasonAVG (the red solid line) or the PS1 individual points (the blue solid line) agree
well in this case. The black dashed line is the proper motion estimate ($-11.22\pm1.85$\masyr) from the PS1 PV3 catalog, combining the PS1 individual points with 2MASS. This proper motion is larger than others, possibly because higher weight is assigned to the 2MASS point for the proper motion fit in PV3 catalog. Among these different proper motion estimates, the fit using seasonAVG positions turned out to be best. It appears accurate and is easy to fit, and we adopt the seasonAVG fit mode for all the stars in the GPS1 construction.

\subsection{Cross-validation of the PS1 Position Uncertainties}\label{sect:diag}
\begin{figure}[!t]
\centering
\includegraphics[scale=0.55]{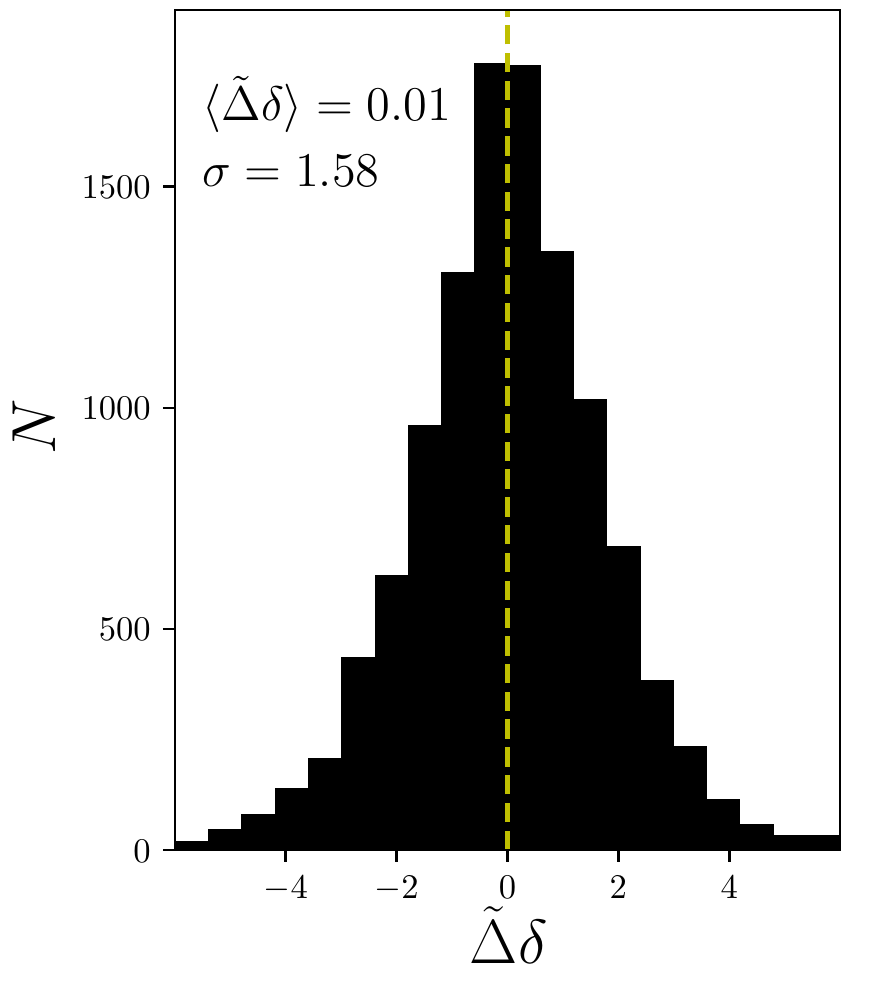}
\caption{The histogram of the normalized residuals obtained from the cross-validation, to check the uncertainties of PS1 SeasonAVG positions. \HJT{The dashed yellow line marks the zero location.}
}
   \label{fig:ddec}
\end{figure}

\HJT{We can use cross-validation to test whether the PS1 SeasonAVG position uncertainties are realistic.
To do so, we randomly choose 2000 bright stars ($14.5<m_r<17.5$) from PS1 and take the 
difference between the observed SeasonAVG PS1 position in a
season, and the value predicted for that season by the proper motion fit
where that particular season has been withheld. Figure \ref{fig:ddec} shows the histogram of the residuals for the sample, normalized with the uncertainty in the PS1 SeasonAVG position, i.e., $\tilde{\Delta}_{\delta}=(\delta_o - \delta_p)/\epsilon_{\delta}$, where $\epsilon_{\delta}$ is the uncertainty of position, $\delta_o$ and $\delta_p$ are the observed and predicted PS1 SeasonAVG positions, respectively.}

Ideally, the width of the histogram should be unity, \HJT{but it is $\sim1.58$.} Tests on mock data have shown that the cross-validation
systematically overestimates the \HJT{deviations}, when the number of data
points in a proper motion fit is small ($<10$ points) in the simulation. For example, even though the
true astrometric uncertainty in a mock sample was set to 5 mas, the
cross-validation analysis returned the uncertainty of 7 mas.

\HJT{Thus, we can conclude that the PS1 SeasonAVG uncertainties are realistic within a factor of 1.5.}

\section{Results and performance}\label{sect:result}

Using the approach described in \HJT{Section \ref{sect:method}}, we determine proper motions for around 350 million stars,
\HJT{to magnitude of $\sim 20$\, in r-band. The catalog draws on PS1 SeasonAVG and Gaia DR1 as the primary data, together with the best available combinations of other surveys. The final catalog uses the robust fit (where all the data points  are fitted regardless of outliers), and cross-validation fit (where outliers are removed while fitting). For reference, we also include a fit without Gaia and one with only PS1 SeasonAVG points.} Table \ref{tab:gps1} lists the main columns contained in the catalog. In the following sub-sections, we discuss the precision and accuracy of proper motions in different cases.

\begin{table*}
 \begin{threeparttable}
\caption{The columns of GPS1 catalog}.\label{tab:gps1}
\centering
\begin{tabular}{l|l|l|l}
\hline
\hline
&Column&Unit&description\\
\hline
1&obj\_id\footnotemark&-&The unique \HJT{but internal} object\_id in PS1\\
2&ra&degree&Right ascension at J2015.0 from Gaia DR1\\
3&dec&degree&Declination at J2015.0 from Gaia DR1\\
4&e\_ra&mas&Positional uncertainty in right ascension at J2015.0 from Gaia DR1\\
5&e\_dec&mas&Positional uncertainty in declination at J2015.0 from Gaia DR1\\
6&ra\_ps1&degree&Average right ascension at J2010 from PS1 PV3\\
7&dec\_ps1&degree&Average declination at J2010 from PS1 PV3\\
8&pmra&\masyr&Proper motion with robust fit in $\alpha\cos\delta$\\
9&pmde&\masyr&Proper motion with robust fit in $\delta$\\
10&e\_pmra&\masyr&Error of the proper motion with robust fit in $\alpha\cos\delta$\\
11&e\_pmde&\masyr&Error of the proper motion with robust fit in $\delta$\\
12&chi2pmra&-&$\chi_{\nu}^2$ from the robust proper motion fit in $\alpha\cos\delta$\\
13&chi2pmde&-&$\chi_{\nu}^2$ from the robust proper motion fit in $\delta$\\
14&pmra\_x&\masyr&Proper motion with cross-validated fit in $\alpha\cos\delta$\\
15&pmde\_x&\masyr&Proper motion with cross-validated fit in $\delta$\\
16&e\_pmra\_x&\masyr&Error of the proper motion with cross-validated fit in $\alpha\cos\delta$\\
17&e\_pmde\_x&\masyr&Error of the proper motion with cross-validated fit in $\delta$\\
18&pmra\_ng&\masyr&Proper motion with no Gaia robust fit in $\alpha\cos\delta$\\
19&pmde\_ng&\masyr&Proper motion with no Gaia robust fit in $\delta$\\
20&e\_pmra\_ng&\masyr&Error of the proper motion with no Gaia robust fit in $\alpha\cos\delta$\\
21&e\_pmde\_ng&\masyr&Error of the proper motion with no Gaia robust fit in $\delta$\\
22&pmra\_ps&\masyr&Proper motion with only PS1 robust fit in $\alpha\cos\delta$\\
23&pmde\_ps&\masyr&Proper motion with only PS1 robust fit in $\delta$\\
24&e\_pmra\_ps&\masyr&Error of the proper motion with only PS1 robust fit in $\alpha\cos\delta$\\
25&e\_mude\_ps&\masyr&Error of the proper motion with only PS1 robust fit in $\delta$\\
26&chi2pmra\_ps&-&$\chi_{\nu}^2$ from only PS1 robust fit in $\alpha\cos\delta$\\
27&chi2pmde\_ps&-&$\chi_{\nu}^2$ from only PS1 robust fit in $\delta$\\
28&n\_obsps1&-&The number of SeasonAVG observations used in the proper motion fit \\
29&n\_obs&-&The number of all the observations used in the robust proper motion fit\\
30&flag\footnotemark&-&An integer number used to flag the different data combination in the proper motion fit. \\
31&magg&mag&g-band magnitude from PS1 \\
32&magr&mag&r-band magnitude from PS1\\
33&magi&mag&i-band magnitude from PS1\\
34&magz&mag&z-band magnitude from PS1\\
35&magy&mag&y-band magnitude from PS1\\
36&e\_magg&mag&Error in g-band magnitude from PS1\\
37&e\_magr&mag&Error in r-band magnitude from PS1\\
38&e\_magi&mag&Error in i-band magnitude from PS1\\
39&e\_magz&mag&Error in z-band magnitude from PS1\\
40&e\_magy&mag&Error in y-band magnitude from PS1\\
41&maggaia&mag&G-band magnitude from Gaia \\
42&e\_maggaia&mag&Error in G-band magnitude from Gaia\\
\hline
\hline
\end{tabular}
 \begin{tablenotes}
  \item [a] \HJT{Here objID is an internal PS1 ID, which is different from the public ID released in PS1 catalog.}
 \item [b] \HJT{In order to label the different survey combinations for proper motion fit, we assign PS1, 2MASS, SDSS, and Gaia with different integer identifiers, i.e. 0, 5, 10, and 20, respectively, and define a $flag$ with the sum of identifiers of surveys combined.}
  \end{tablenotes}
 \end{threeparttable}
\end{table*}

\subsection{Uncertainties in Proper Motions}

The \HJT{footprint overlap} among Gaia, PS1, SDSS and 2MASS surveys introduces some complexity: $\sim$ 23\% stars are covered by Gaia, PS1, and SDSS, $\sim$ 73\% by PS1 and Gaia, but not SDSS, $\sim$ 3\% stars are only observed by PS1. Therefore, it is necessary to investigate how the final proper motions are affected by combining different data sets. 

\begin{table*}
 \begin{threeparttable}[b]
\caption{The formal fitting uncertainties of the proper motions in the different data combinations}.\label{tab:dif_comb}
\centering
\begin{tabular}{c|l|c|c|c|c|c|c|c}
\hline
\hline
ID&Mode& \multicolumn{2}{c|}{$m_r<14$} &\multicolumn{2}{c|}{$14<m_r<18$\tnote{a}}&\multicolumn{2}{c|}{$m_r>18$} &{$N$\tnote{b}} \\
\hline
&&\mmura &\mmudec &\mmura &\mmudec &\mmura &\mmudec & \\
\hline
\multicolumn{2}{c|}{}&\multicolumn{6}{c|}{\masyr}&\\
\hline
1&GPS (Gaia+PS1+SDSS+2MASS)&1.74$\pm$0.54&1.53$\pm$0.46&1.35$\pm$0.33&1.21$\pm$0.29&1.89$\pm$0.51&1.72$\pm$0.48&50000\\
2&GP  (Gaia+PS1+2MASS)&1.67$\pm$0.52&1.45$\pm$0.51&1.41$\pm$0.41&1.23$\pm$0.35&2.59$\pm$1.11&2.20$\pm$0.96&50000\\
3&PD  (PS1+SDSS+2MASS)&3.00$\pm$1.01&2.60$\pm$1.02&1.91$\pm$0.48&1.72$\pm$0.45&2.65$\pm$0.85&2.44$\pm$0.79&50000\\
4&PS1 ({\sl only} PS1)&4.34$\pm$2.55&3.51$\pm$2.05&2.53$\pm$0.90&2.12$\pm$0.73&4.58$\pm$2.33&3.75$\pm$1.88&50000\\
\hline
\hline
\end{tabular}
 \begin{tablenotes}
 \item [a] The uncertainties of the proper motions in $14<m_r<18$ are marked by black dashed lines in Figure \ref{fig:sigmu_mr_GPS1}.
 \item [b] We randomly select 50,000 objects from across the sky to derive these statistics. The proper motion estimates from the GPS, PS, and PS1 modes are provided for each star in the GPS1 catalog. So, the analysis for these three cases is based on the same sample taken from within the SDSS. Only the objects in the non-SDSS regions have proper motion estimates using only GP. Therefore, the statistics for the GP mode in the table  are derived from a sample outside the SDSS coverage.
  \end{tablenotes}
 \end{threeparttable}
\end{table*}

\subsubsection{The Different Data Set Combinations}\label{sect:all_fitting}
We now investigate how the uncertainties in proper motion \HJT{differ among} the following four combinations of data sets:  Gaia + PS1 + SDSS + 2MASS (GPS), Gaia + PS1 + 2MASS (GP), PS1 + SDSS + 2MASS (PD), and 
{\sl only} PS1 (PS1). \HJT{For the catalog table, different} surveys are assigned different integer \HJT{identifiers: 0, 5, 10, and 20 for PS1, 2MASS, SDSS, and Gaia, respectively. This defines a $flag$ for different survey combinations entering a fit}, represented as the sum of \HJT{the individual survey identifiers}. The primary observations are those from PS1, so the positions for each star must include the PS1 detections when fitting for proper motion.

\begin{figure*}[!t]
\centering
\includegraphics[width=0.504\textwidth, trim=0.0cm 0.0cm 0.1cm 0.0cm, clip]{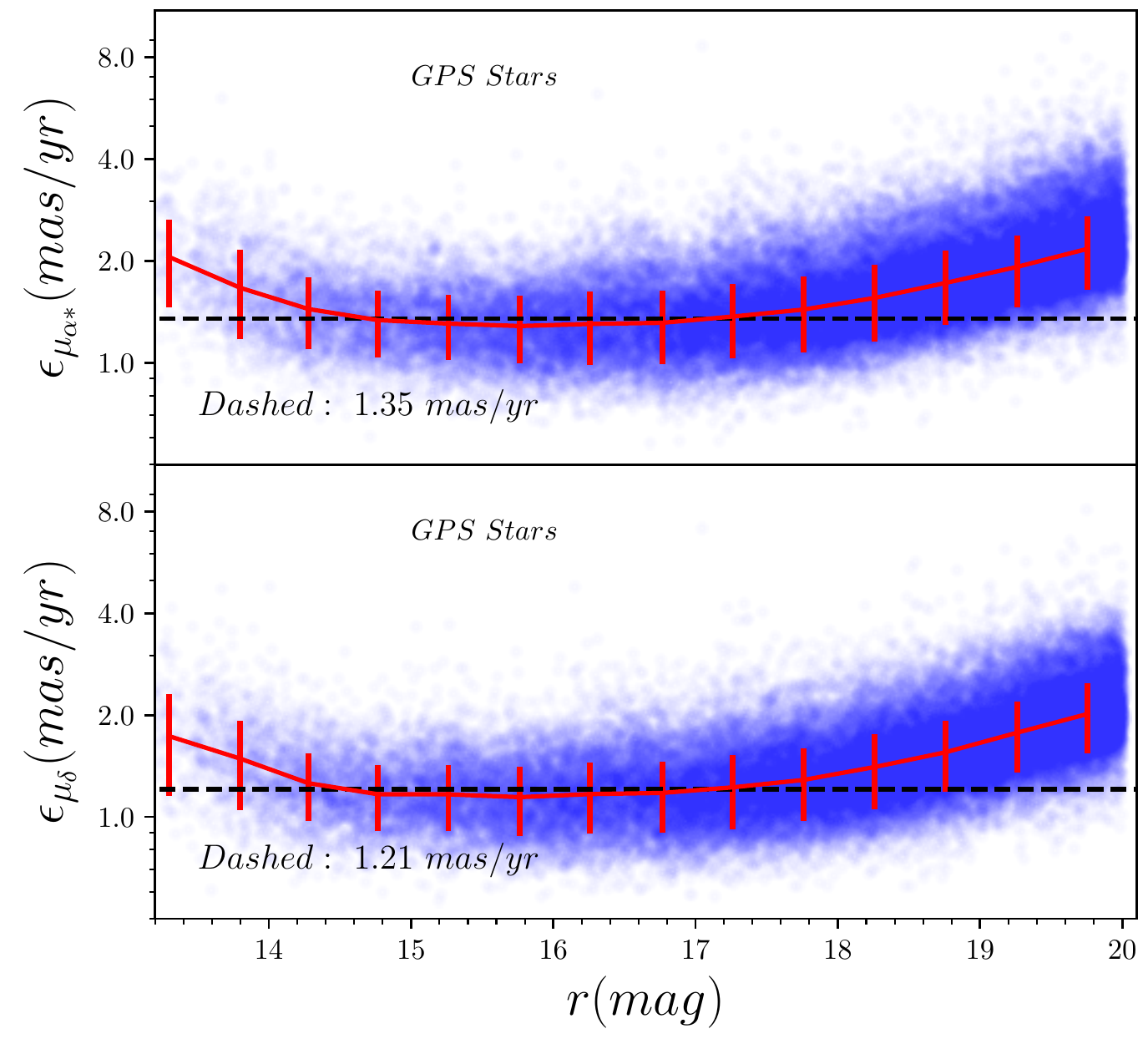} 
\includegraphics[width=0.44\textwidth, trim=1.85cm 0.0cm 0.0cm 0.0cm, clip]{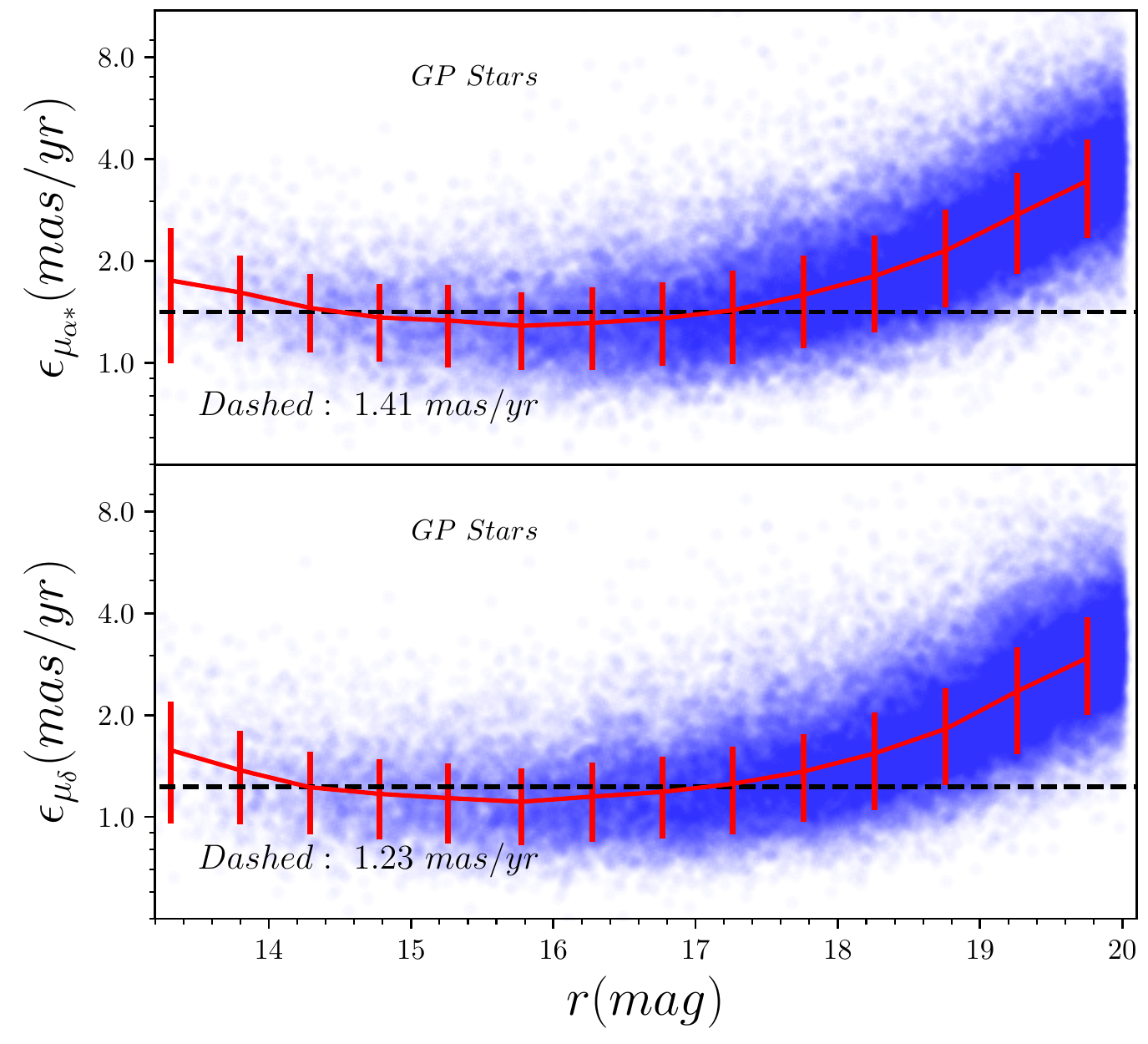} 
\includegraphics[width=0.504\textwidth, trim=0.0cm 0.0cm 0.1cm 0.0cm, clip]{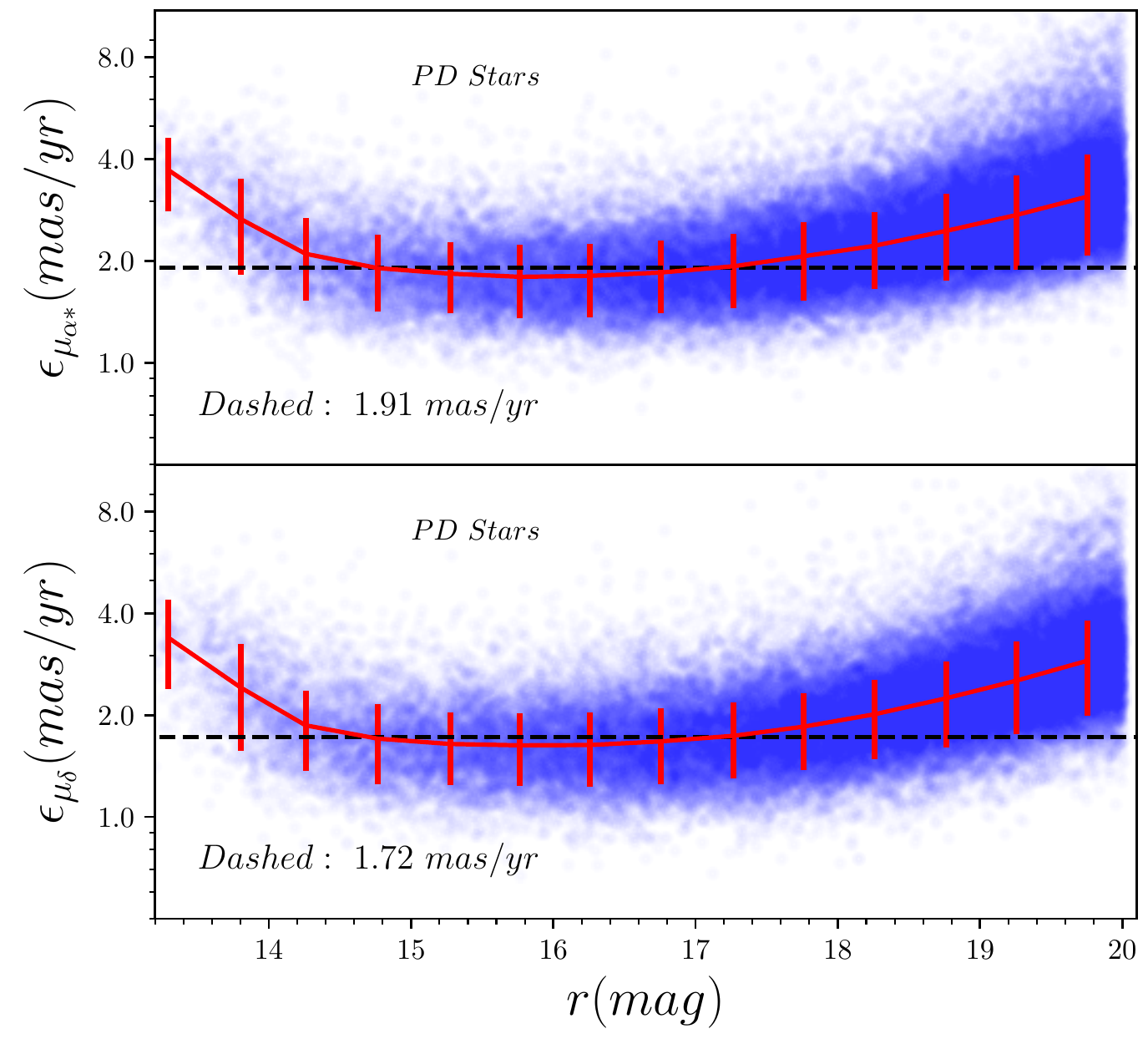} 
\includegraphics[width=0.44\textwidth, trim=1.85cm 0.0cm 0.0cm 0.0cm, clip]{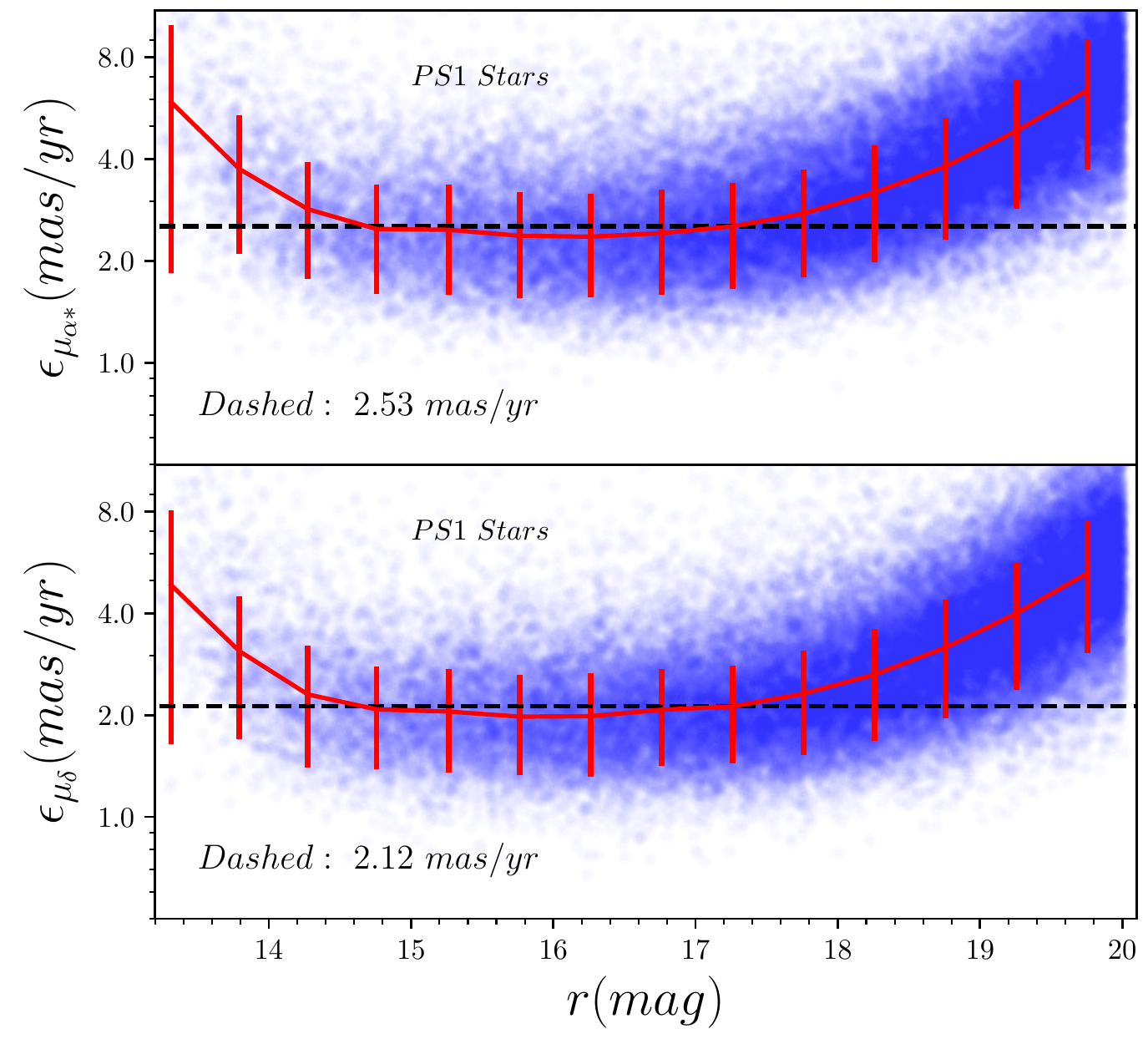} 
\caption{Proper motions precision for the four different combinations of data sets (top-left: GPS, top-right: GP, bottom-left: PS, and bottom-right: ONLY PS1). In the four panels, the red curves are the median uncertainties of proper motions within different magnitude bins, and the black lines mark typical average uncertainties in the magnitude range  $14<m_r<18$. The blue scatter points represent the million stars randomly selected from the sky. All the uncertainties are logarithmic in every y-axis.
The typical average uncertainties for the four combination modes are \dmura\ $\sim$1.35 \masyr, \dmudec\ $\sim$1.21 \masyr\ for the GPS mode (top-left), \dmura\ $\sim$1.41 \masyr, \dmudec\ $\sim$1.23 \masyr\ for the GP mode (top-right), \dmura\ $\sim$1.91 \masyr, \dmudec\ $\sim$1.72 \masyr\ for the PD mode (bottom-left), and \dmura\ $\sim$2.53 \masyr, \dmudec\ $\sim$2.12 \masyr\ for the ONLY PS1 mode (the bottom-right), respectively. 
}
   \label{fig:sigmu_mr_GPS1}
\end{figure*}

Figure \ref{fig:sigmu_mr_GPS1} summarizes the distribution of proper motion uncertainties for the four different combinations. In the four panels, the blue points correspond to the 50,000 stars randomly selected from the sky and the red curves are the median uncertainties in proper motions within different magnitude bins. \HJT{The average uncertainties in magnitude bins are listed in Table \ref{tab:dif_comb}, with the mean ($14<m_r<18$) marked by black lines. In the GPS mode (see Table \ref{tab:dif_comb}), the average uncertainties are \dmura\ $\sim$1.35 \masyr\ and \dmudec\ $\sim$1.21 \masyr. This is slightly better than the GP mode (\mura\ $\sim$1.41 \masyr and \mudec\ $\sim$1.23 \masyr). Without Gaia positions (PD mode), the typical uncertainties become \dmura\ $\sim$1.91 \masyr and \dmudec\ $\sim$1.72 \masyr. Gaia positions improve the precision by $\sim0.6$\masyr\ for both the \dmura\ and \dmudec. For PS1 data alone, the mean uncertainties become \dmura\ $\sim$2.53 \masyr and \dmudec\ $\sim$2.12 \masyr. The precision improvement is dominated by Gaia ($\sim1.0$\masyr\ on an average).}  

For the brighter stars with $m_r<14$, the proper motion uncertainties
increase as the PS1 detections begin to saturate. A comparison of the GP mode with the PS1 mode reveals that Gaia can improve the precision of the bright stars by $\sim2.1$\masyr, as shown in Table \ref{tab:dif_comb}. \HJT{Comparing with the PS1 mode, we find that SDSS in the PD mode} can improve the uncertainty by $\sim1.1$\masyr. Therefore, Gaia detections are \HJT{also} more effective in reducing uncertainties than SDSS for the case of bright stars with $m_r<14$.

For the fainter stars with $m_r>18$, \HJT{the positional uncertainties are worse than those of brighter stars}, implying that the precision of the obtained proper motions will be worse towards the faint end. As the values in Table \ref{tab:dif_comb} show, both SDSS and Gaia can improve the precision of the proper motions in the PS1 mode by $\sim1.6$\masyr\ individually, and by $\sim2.0$\masyr\ together. Therefore, Gaia and SDSS are comparably important for reducing uncertainties \HJT{for the faint stars}.

\begin{figure*}[!t]
\centering
\includegraphics[width=0.45\textwidth, trim=0.0cm 0.0cm 0.0cm 0.0cm, clip]{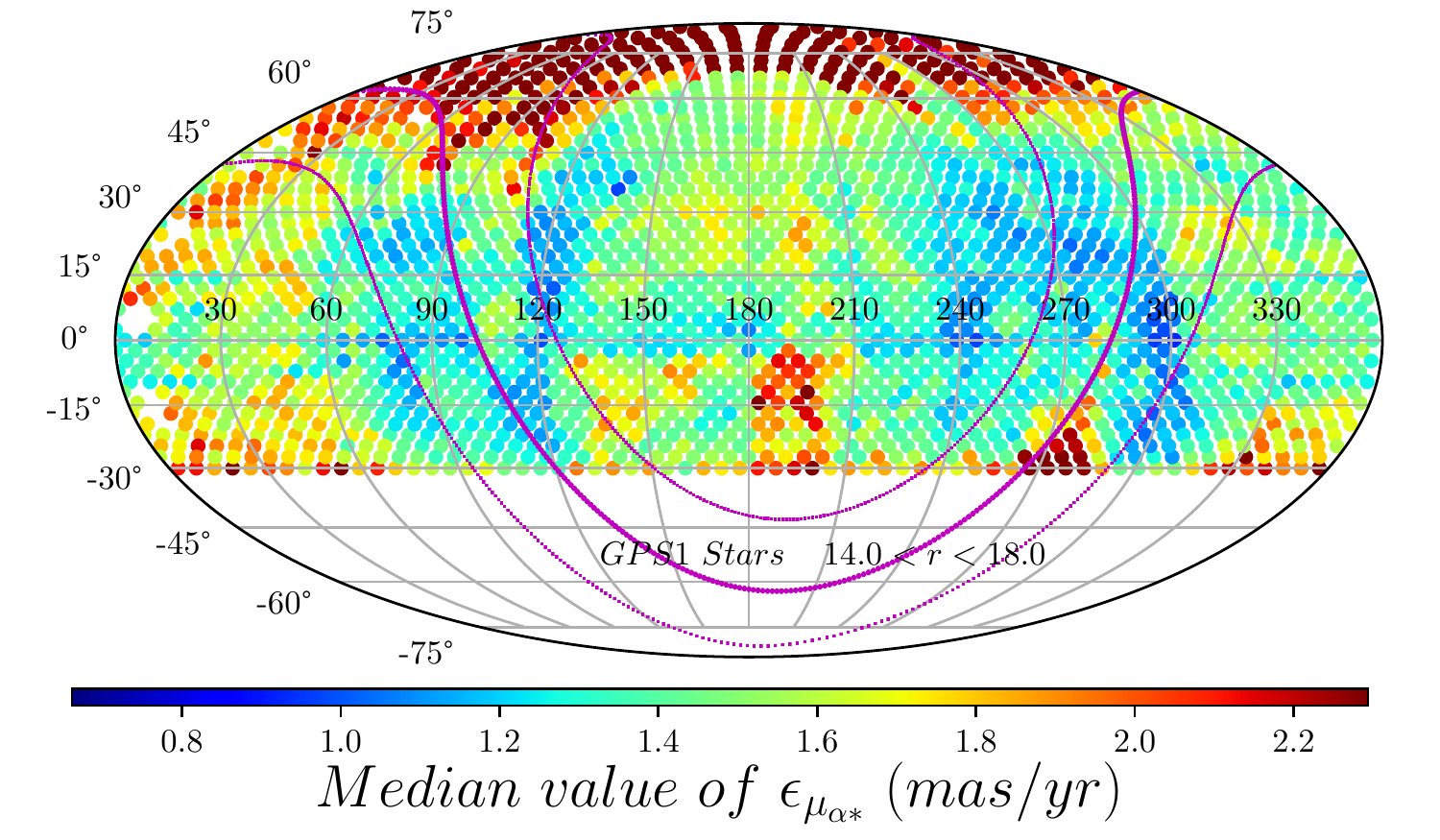}
\includegraphics[width=0.45\textwidth, trim=0.0cm 0.0cm 0.0cm 0.0cm, clip]{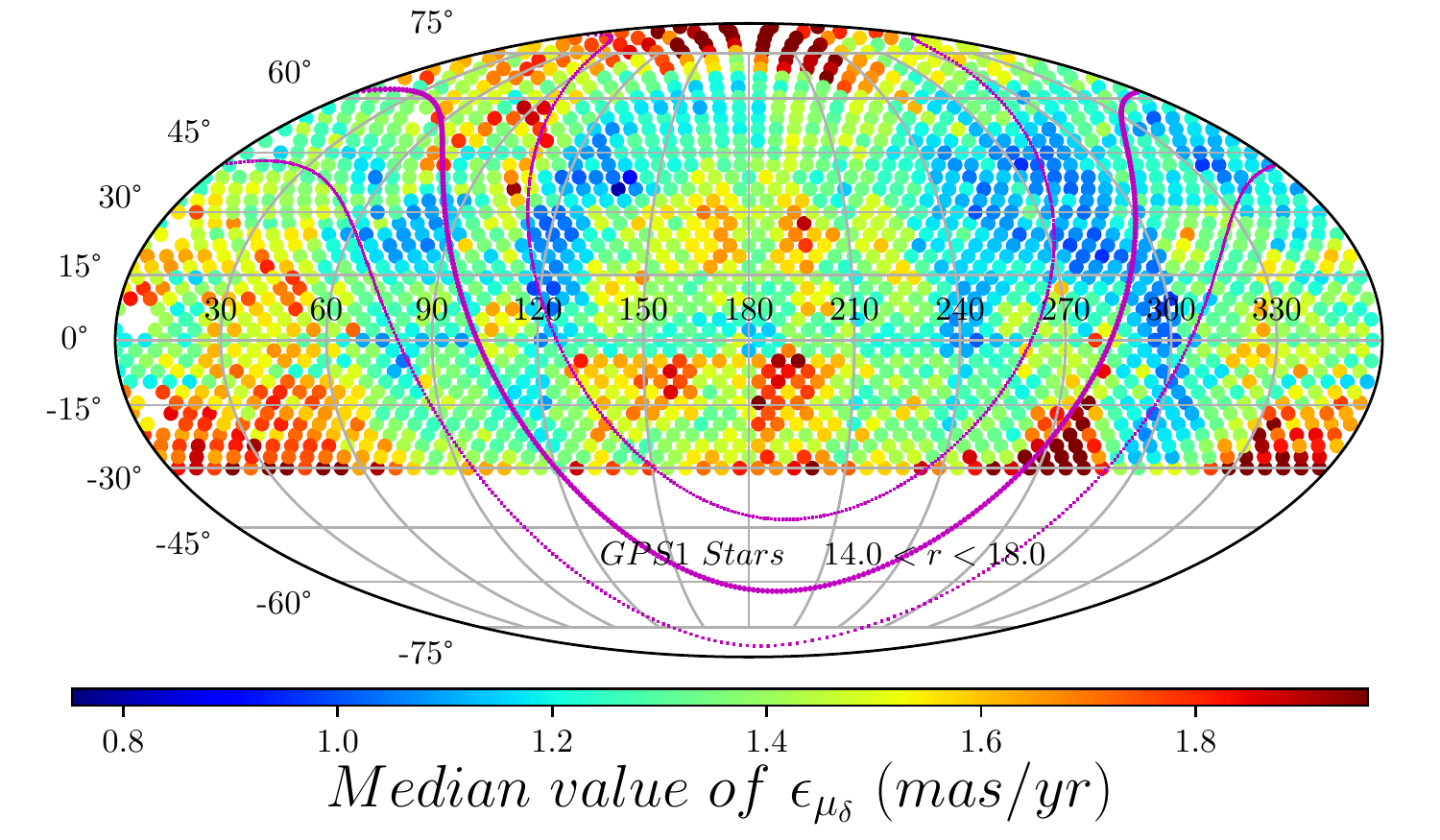}
\caption{The distribution of proper motion uncertainties for stars with $14<m_r<18$; this is illustrated with an equatorial Mollweide projection of the entire 3$\pi$ sky region. 
The pink solid ($b=0^{\circ}$) and two dotted lines ($b=\pm20^{\circ}$) mark the location of the Galactic plane in the equatorial coordinate system, \HJT{where sources are crowded and the effects of dust extinction are manifest \citep{tian2014}}. 
To highlight the structures in the maps, the color bar is scaled in $\pm3\sigma$ around the entire median value for each map. 
}\label{fig:uncertanties_star}
\end{figure*}

For stars with $14<m_r<18$, the GPS1 catalog is at its best. Photon noise matters little, yet the sources are not saturated. Figure \ref{fig:uncertanties_star} illustrates the distribution of uncertainties of these stars as Mollweide projection maps of the entire 3$\pi$ region of the sky in equatorial coordinate system, containing six million stars randomly selected. The median uncertainty in each pixel is calculated \HJT{from hundreds of stars}.
The median values of the uncertainties are $\sim1.5$\masyr\ for both \mura\ (the left panel) and \mudec\ (the right panel), as shown in the maps. The uncertainties at high and low declinations are larger, \HJT{as SDSS data missing}. The small uncertainties in the north Galactic cap are driven by the SDSS observations taken ten or fifteen years ago. 

\begin{figure}[!t]
\centering
\includegraphics[width=0.46\textwidth, trim=0.0cm 0.0cm 0.0cm 0.0cm, clip]{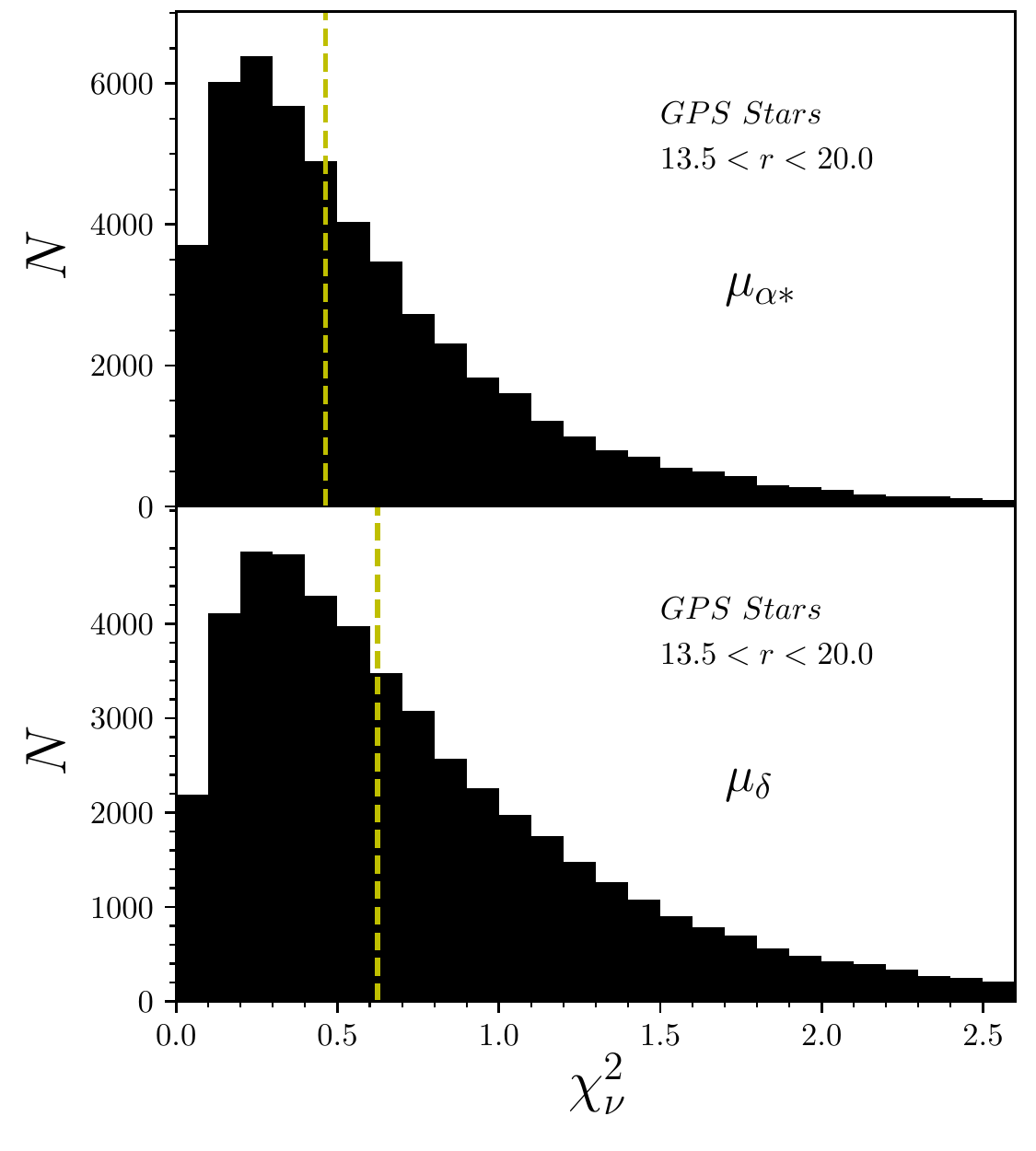}
\caption{Illustration of the quality of the proper motion fits. Shown is the reduced $\chi^2$ of the fits for the GPS case. The median values for both \mura\ and \mudec\ (marked with the yellow dashed lines) are smaller than 1, implying that most fits are good, and that the individual epochs' uncertainties are (slightly) conservative estimates.
}\label{fig:hist_chi2}
\end{figure}

Figure \ref{fig:hist_chi2} shows the distribution of reduced $\chi^2$ for the proper motion fit \HJT{for a random subset of stars.} This suggests that the proper motions are well fit for most of the stars, and that perhaps the individual uncertainty estimates are \HJT{somewhat} conservative. \HJT{The actual uncertainties may be slightly smaller than our estimates}.


\subsection{Validation of Proper Motions}\label{sec:val}

We now turn to the astrophysical validation of the derived
proper motions, using galaxies, QSOs, distant stars and star clusters with well-known proper motions. All these validations have issues that require attention:
galaxies and QSOs are distant enough to know {\it a priori} that their proper motions can be neglected; but galaxies are extended and often asymmetric objects, and QSOs with their strong emission lines show peculiar differential chromatic refraction (DCR). Stars in the Galactic halo are simple point sources, but may not be distant enough
to \HJT{have negligible} proper motion: in particular, reflex of the Sun's motion is still observable up to $30$ kpc. Member stars of open clusters share a common motion, but non-member contamination may be difficult to remove. Sources bright enough to have TGAS proper motions are too bright to be in the present sample. Therefore, there is no simple, ideal set of astrophysical sources to easily validate our proper motion estimates. 

\subsubsection{Validation with Galaxies}\label{sect:val_gal}

\begin{figure*}[!t]
\centering
\includegraphics[width=0.45\textwidth, trim=0.0cm 0.0cm 0.0cm 0.0cm, clip]{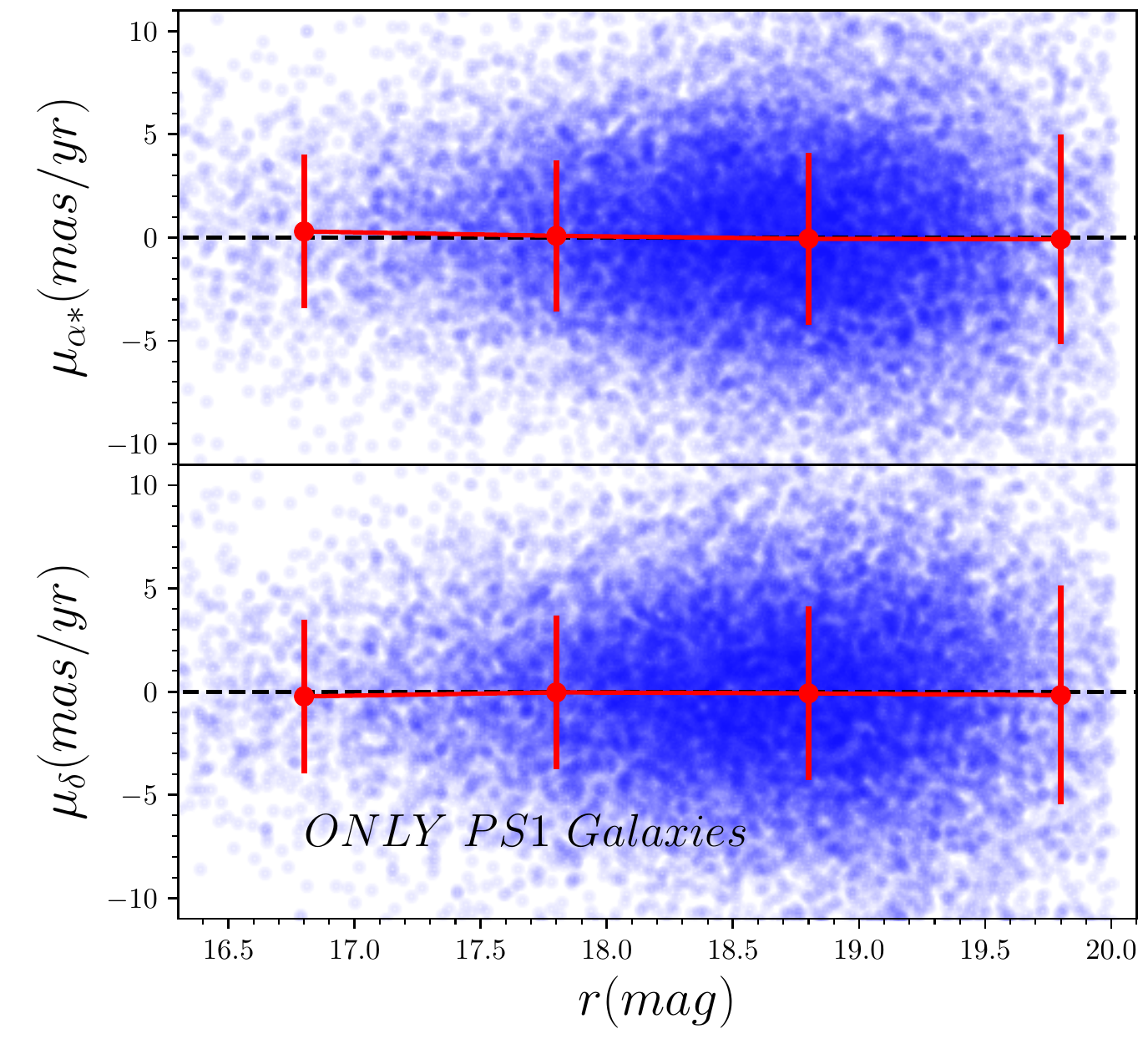}
\includegraphics[width=0.45\textwidth, trim=0.0cm 0.0cm 0.0cm 0.0cm, clip]{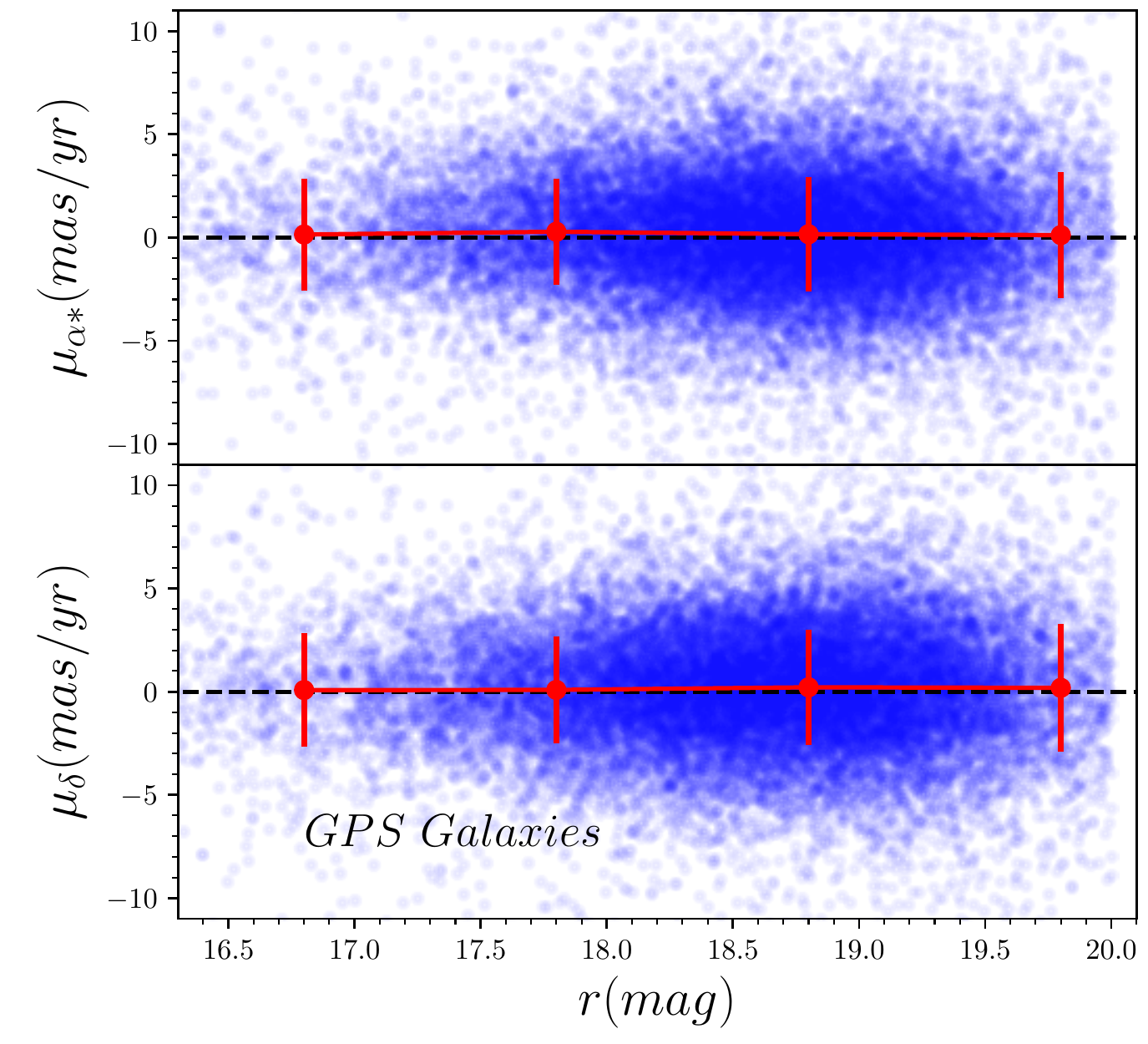}
\caption{Validation of proper motions with galaxies in different magnitude bins. Two typical proper motions are presented: the PS1 proper motions (the left panel), and the GPS proper motions (the right panel). The red curves are the median values of proper motion in different magnitude bins and the error bars represent the robust {\it rms}. The black dashed lines mark zero proper motion. All the red points oscillate around the black dashed lines  within $\pm$0.3 \masyr\, indicating that the accuracy of the proper motion is better than 0.3 \masyr. 
The average {\it rms} in the ONLY PS1 case is $\sim$ 4.2 \masyr, which is reduced to $\sim$2.8 \masyr\ in the GPS case.}\label{fig:vali_gals}
\end{figure*}

We select a sample of galaxies from the region covered by the PS1, SDSS, and Gaia surveys, and calculate the proper motions in two data combinations: PS1 and GPS modes. Figure \ref{fig:vali_gals} shows \HJT{that the median of these apparent (and presumably spurious) proper motions lies within $\pm$0.3 \masyr\ of zero, implying that the accuracy of proper motion is better than 0.3 \masyr. It also implies that this is a consistency check, since we used galaxies to build the reference frame, the proper motions of galaxies should be zero by design. The actual precision for galaxies is of course worse than that for stars, as they are extended.} 

\subsubsection{Validation with QSOs}\label{sect:val_qso}

\citet{nina2016} identified a sample of \HJT{over a} million QSO candidates from the PS1 3$\pi$ survey image data. QSOs have strong emission lines, which cause subtle image centroid effects,
when differential chromatic refraction (DCR) comes into play. For validation, we only choose QSOs \HJT{with high probability} in $14.0<m_r<17.5$. 

\begin{figure*}[!t]
\centering
\includegraphics[width=0.45\textwidth, trim=0.0cm 0.0cm 0.0cm 0.0cm, clip]{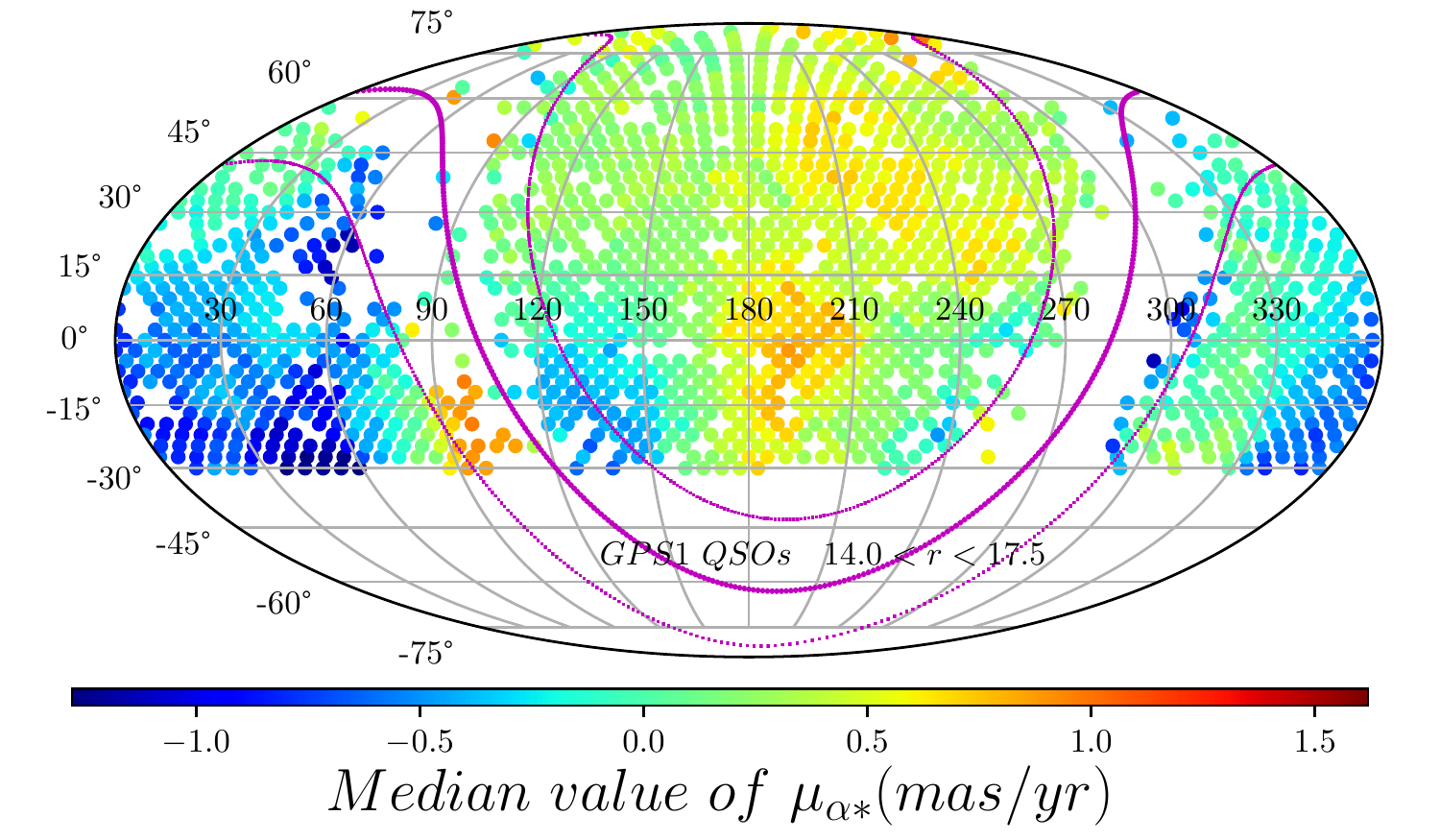}
\includegraphics[width=0.45\textwidth, trim=0.0cm 0.0cm 0.0cm 0.0cm, clip]{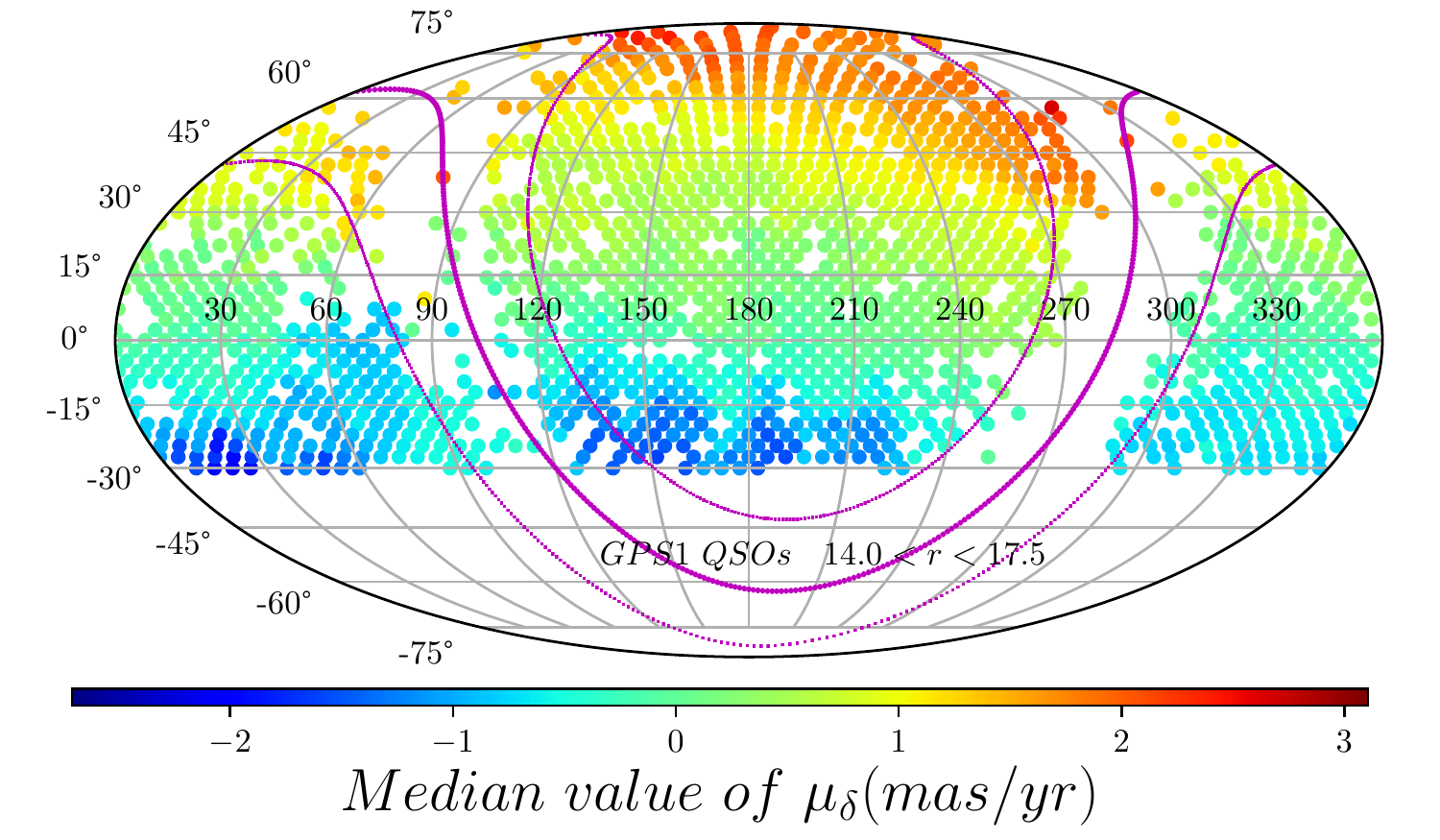}
\caption{Validation of proper motions with QSOs in Mollweide projection map of the entire 3$\pi$ sky region in the equatorial coordinate system. 
The pink solid ($b=0^{\circ}$) and two dotted lines ($b=\pm20^{\circ}$) represent the location of the Galactic plane in the equatorial coordinate system. 
}\label{fig:vali_qso0}
\end{figure*}

\begin{figure*}[!t]
\centering
\includegraphics[width=0.45\textwidth, trim=0.0cm 0.0cm 0.0cm 0.0cm, clip]{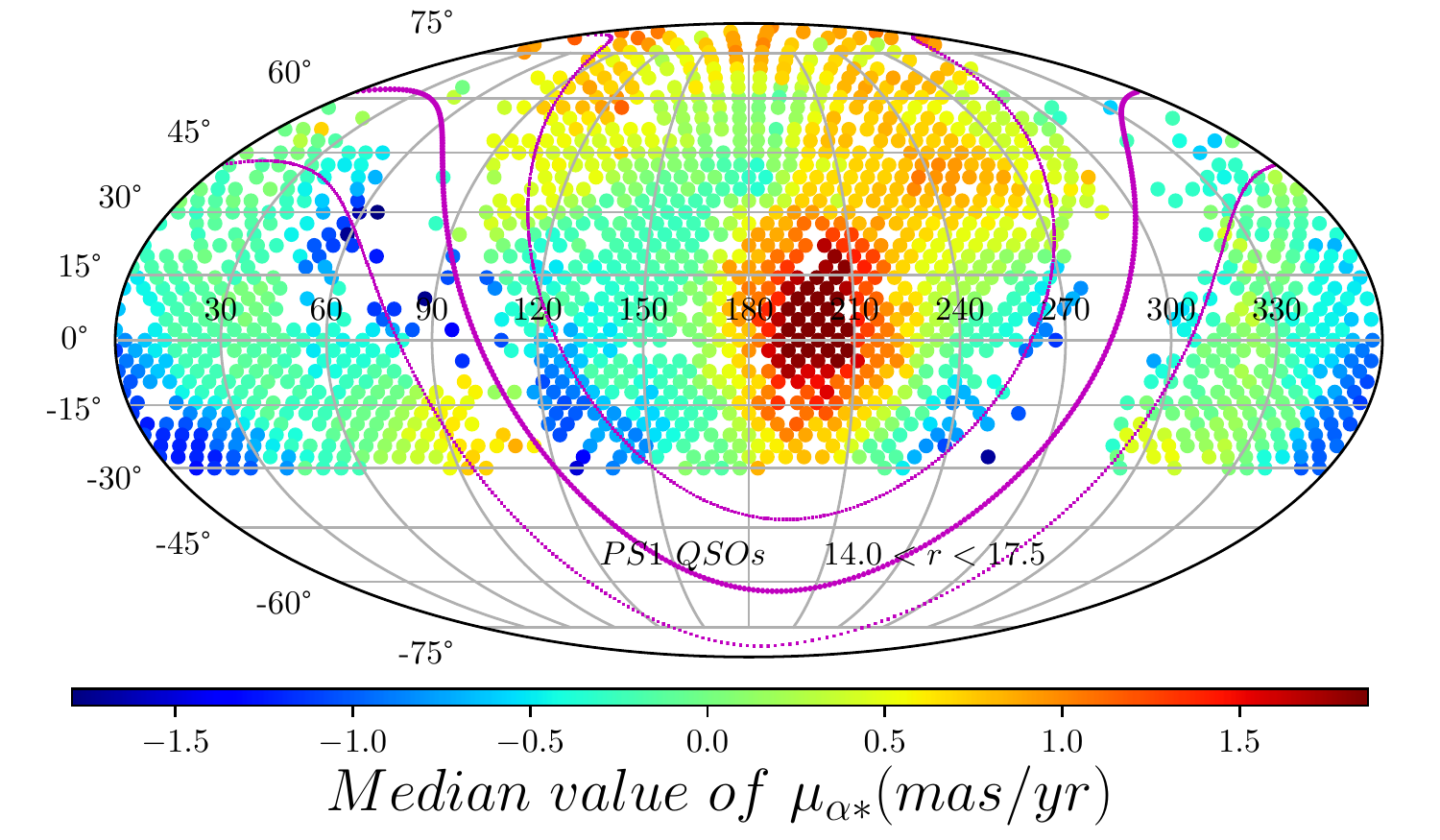}
\includegraphics[width=0.45\textwidth, trim=0.0cm 0.0cm 0.0cm 0.0cm, clip]{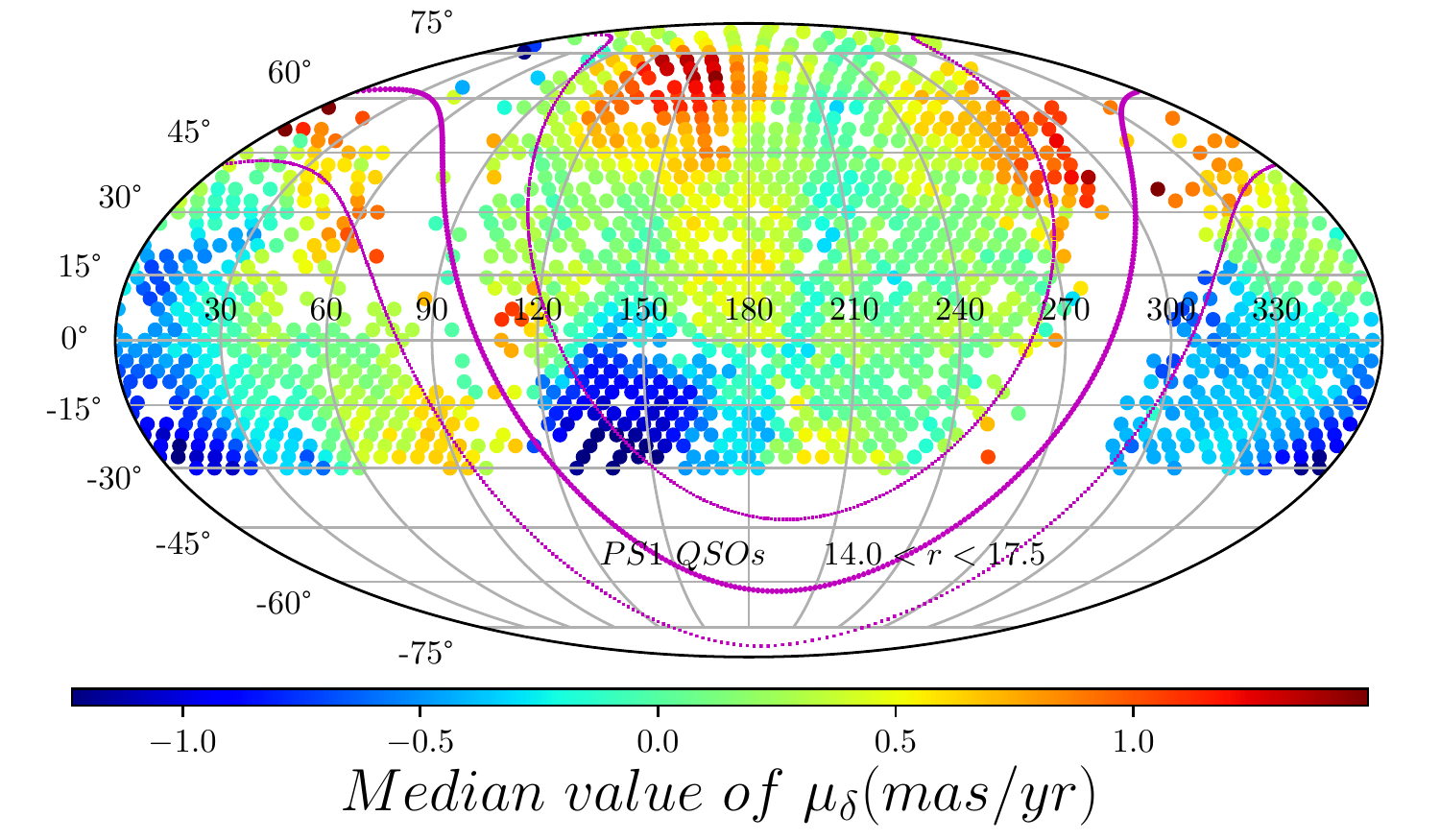}
\caption{Same as Figure \ref{fig:vali_qso0}, but the proper motions of the QSOs in the maps are measured when only fitting PS1 SeasonAVG points.
}\label{fig:vali_qso1}
\end{figure*}

Figure \ref{fig:vali_qso0} displays the apparent (and spurious, if significantly non-zero) proper motion distributions of the QSOs, showing the entire PS1 3$\pi$ sky region
in an equatorial Mollweide projection. The median value in each pixel is calculated using QSOs that lie within a radius of $10^{\circ}$, a size that ensures inclusion of at least tens of QSOs. The apparent proper motions of QSOs show a significant non-zero pattern across the sky \HJT{especially in $\delta$}. At high declinations, the $\delta$ proper motions are biased \HJT{by} up to 2\masyr. At low declinations, the $\delta$ proper motions are slightly under-estimated by $\sim$ 0.5\masyr. For comparison, Figure \ref{fig:vali_qso1} displays the \HJT{apparent} proper motions measured when only fitting the PS1 SeasonAVG points. Similar to Figure \ref{fig:vali_qso0}, the $\delta$ proper motions (the right panel) in high and low declinations are also over- and under-estimated by an average of $\sim$ 0.5\masyr, respectively. There is an obvious pattern in the region of $170^{\circ}<\alpha<220^{\circ}$ and $-20^{\circ}<\delta<20^{\circ}$ at the map of $\alpha$ proper motions (the left panel). \HJT{It is probably caused by the PS1 observation, since the pattern looks even clearer than that in the GPS1 proper motion map (the left panel in Figure \ref{fig:vali_qso0})}. 
\HJT{We will discuss the bias induced by DCR in Section \ref{sect:dcr}.}


\subsubsection{Validation with Star Clusters}\label{sect:val_gc}
M67 is a well known open cluster with a distance of $\sim$ 850\ pc. Given its well-defined main sequence track, member candidates of the cluster can be easily identified using a color-magnitude diagram (CMD). All member stars should have mutually indistinguishable proper motions, as the cluster has an internal velocity dispersion of $\sim1$~\kms \citep{Geller2015}. Because we know the absolute proper motions of a few M67 members from TGAS \citep{gaia2016a}, M67 could be an ideal testbed for our proper motion {\it accuracy}. But this requires careful accounting of field star contamination. 

\begin{figure}[!t]
\centering
\includegraphics[width=0.40\textwidth, trim=0.25cm 0.3cm 0.0cm 0.62cm, clip]{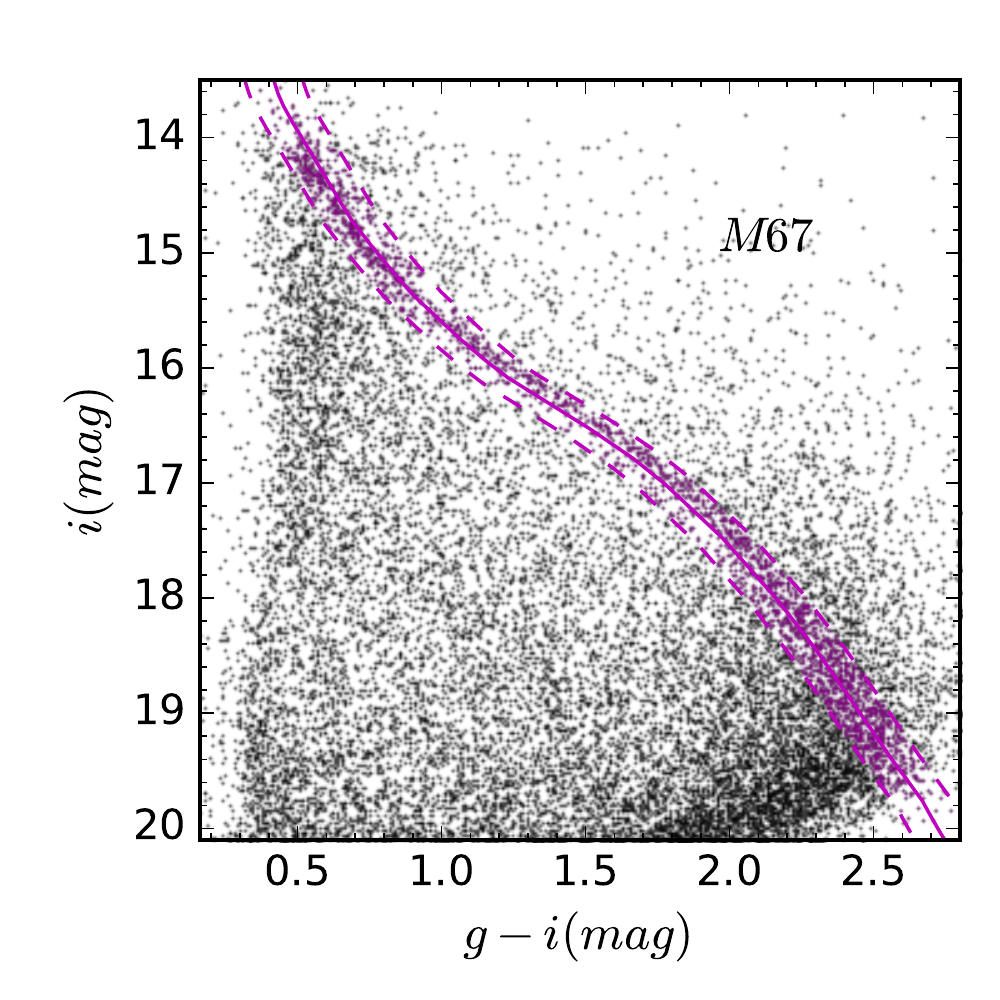}
\caption{The color ($g-i$) and magnitude (i-band) diagram (CMD) \HJT{of stars in the view field of M67}, based on PS1 photometry. The pink solid curve in the CMD is the PARSEC synthetic stellar track, while the two dashed lines, offset by $\pm$0.1 mag serve for our photometric membership definition. 
}
   \label{fig:CMD_M67}
\end{figure}

Figure \ref{fig:CMD_M67} presents a color ($g-i$) and magnitude ($i$-band) diagram,
based on PS1 photometry \HJT{of stars} within an angular radius $r_2=1.03\arcdeg$ of M67 \citep[see][]{Kharchenko2012}. The solid pink curve corresponds to the PARSEC synthetic stellar track built with the Padova web-server CMD 2.8\footnote[1]{$http://stev.oapd.inaf.it/cgi-bin/cmd_2.8$}, while the two dashed curves \HJT{offset by} $\pm$0.1 mag \HJT{define} the region we use to select likely members. We select member candidates by three criteria: (1) $r<r_2$; (2) distributed between the two dashed lines in the CMD (see Figure \ref{fig:CMD_M67}); (3) $13.5<m_r<18.0$.

\begin{figure}[!t]
\centering
\includegraphics[width=0.35\textwidth, trim=0.1cm 0.1cm 0.1cm 0.1cm, clip]{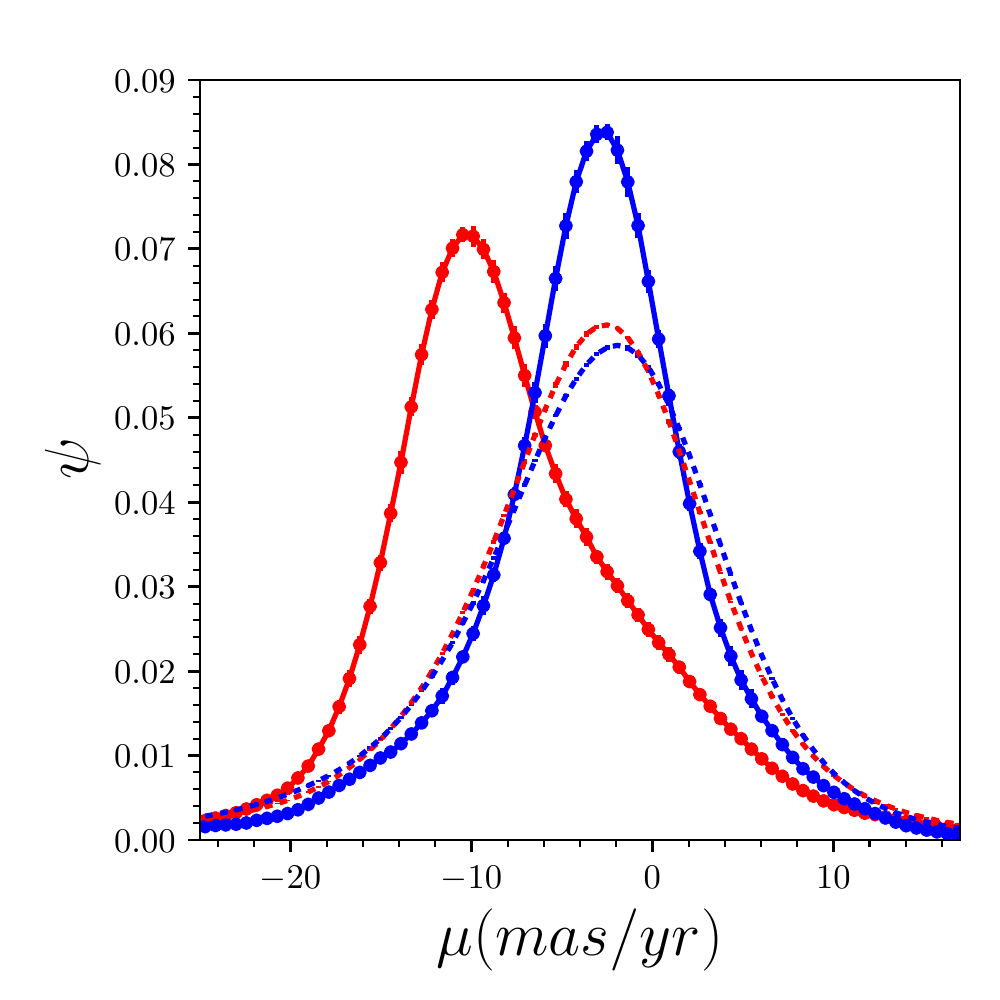}
\caption{The normalized Gaussian-Kernel-Smoothed probability distribution of the proper motions (red for \mura, blue for \mudec) of the stars nearby M67. The solid curves are the distributions ($\psi_{m+f}$) of the member candidates which are composed by both the member and field stars within $r<r_2$. The dashed curves are the distribution($\psi_{f}$) of the field stars within $r_2<r<2r_2$. The error bars are obtained from the 100 bootstrap sub-samples.
}
   \label{fig:pdf_m67}
\end{figure}

Most member stars are located within $r_2$, but field stars might still significantly contaminate the membership in this region ($r<r_2$). Outside $r_2$, the distribution is probably dominated by field contamination. Figure \ref{fig:pdf_m67} shows the normalized Gaussian-kernel-Smoothed probability distribution of proper motions of the stars \HJT{in the field of} M67. The size of the Kernel \HJT{was chosen to match} the precision of the proper motions (2\masyr). 
The solid curves show the distribution ($\psi_{m+f}$) of the member candidates \HJT{selected according to CMD}, which are composed of both member and field stars. The dashed curves show the distribution ($\psi_{f}$) of stars within $r_2<r<2r_2$, which is dominated by field stars. The error bars are obtained using 100 bootstrap sub-samples.

To estimate the \HJT{mean} proper motion of the star cluster, 
we use Markov Chain Monte Carlo (MCMC) simulation\footnote[2]{We use the \emph{emcee} code to run the MCMC~\citep{forman2013}} to determine the most likely values of the five parameters:  (\meanmura, \meanmudec, \smura, \smudec) $=$ (-10.54, -2.94, 3.16, 3.37)\masyr, and $r_m = 54\%$, \HJT{where $r_m$ is the member fraction of all stars within $r_2$}.


\begin{figure}[!t]
\centering
\includegraphics[width=0.40\textwidth, trim=0.25cm 0.3cm 0.0cm 0.62cm, clip]{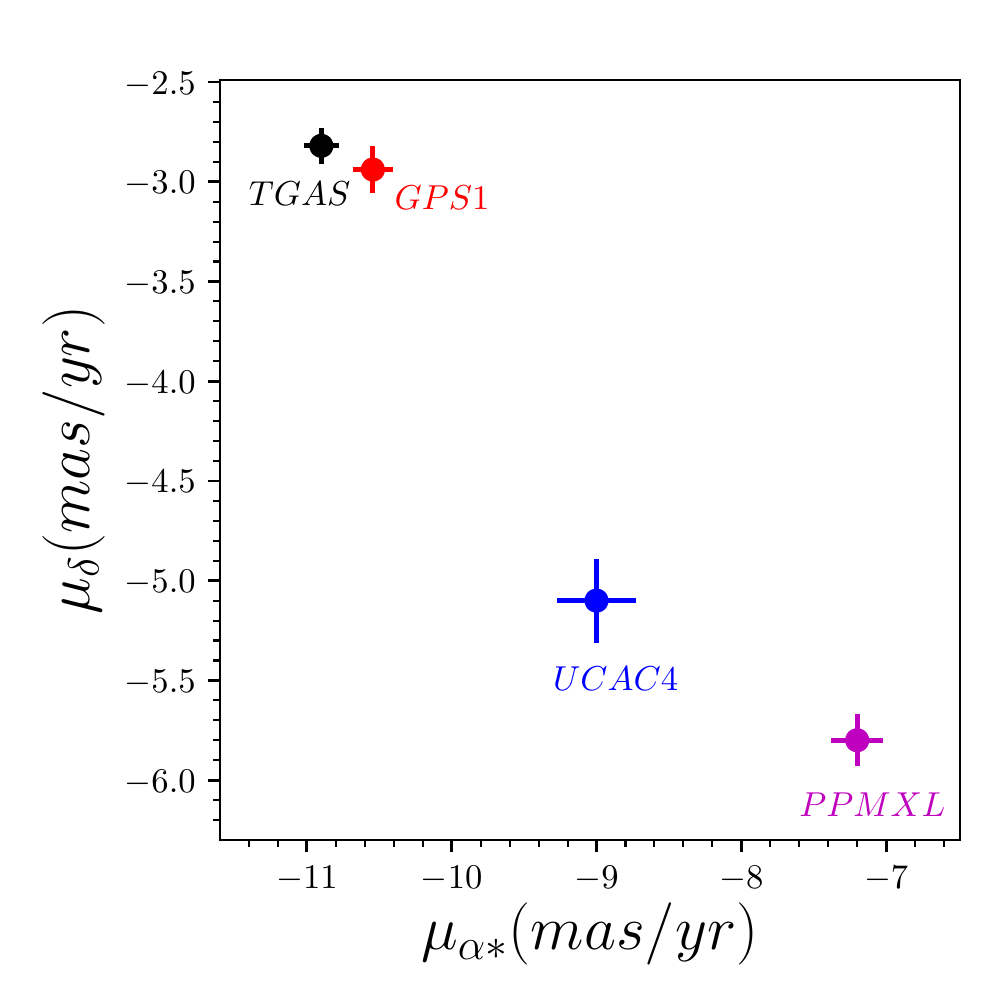}
\caption{Validation of the GPS1 proper motion precision and accuracy, using the open cluster M67.
The black point marks the median proper motion of the two member stars listed in Table \ref{tab:tgas}.
The red point is the mean proper motion obtained from MCMC. 
For comparison, the values from PPMXL (the pink point) and UCAC4 (the blue point) are also displayed in the panel.
}
   \label{fig:vali_M67}
\end{figure}

Within the angular radius of M67, we select two TGAS stars (see Table \ref{tab:tgas}), whose proper motions \HJT{are measured with high precision and the values are consistent with each other within errors. Also, their proper motions and parallaxes match those derived by \citet{bellini2010}. Therefore, their mean proper motion (-10.90$\pm$0.12, -2.82$\pm$0.09 \masyr) can be considered as a robust estimate of the proper motion of M67. For comparison, the mean GPS1 proper motion of the likely cluster members, (\mmura, \mmudec) $=$  (-10.54$\pm$ 0.14, -2.94$\pm$0.13) \masyr\ obtained from MCMC, is remarkably consistent with the robust estimate, as shown in Figure \ref{fig:vali_M67}. This suggests that the GPS1 proper motions are measured not only with a small random error, but also with a tiny systematic error.}

We also compared our M67 proper motions with proper motions provided by PPMXL \citep{roeser2010} and UCAC4 \citep{zacharias2013} catalogs. Table \ref{tab:pm_M67} gives the proper motions of M67 estimated from four different catalogs. The value from TGAS is \HJT{the robust estimate of the proper motion of M67, i.e., the average proper motion of two typical member stars listed in Table \ref{tab:tgas}}. Comparing with the \HJT{robust} value, the proper motion of GPS1 obtained from MCMC simulations shows a systematic offset of $<0.3$ \masyr, almost 10 times better than PPMXL and UCAC4 ($>2.0$ \masyr). Combining with Gaia DR1 data, \citet{Zacharias2017} and \citet{Altmann2017} recently updated the UCAC4 and PPMXL catalogs, and named the new catalogs UCAC5 and HSOY, respectively. Although the precisions of proper motions in the new catalogs are claimed to be improved to 1-5 \masyr\ with one year positions from Gaia DR1, the accuracies are not reported definitely in their papers. Fortunately, UCAC5 mainly focuses on the proper motions of bright sources ($r<15$\, mag) with a precision of $<$2 \masyr, which will fill up the gap in GPS1 catalog.

\begin{table}
\caption{ TGAS Proper Motions for M67 Member Stars}.\label{tab:tgas}
\centering
\begin{tabular}{c|c|c|c|c|c|c}
\hline
\hline
ID&$\alpha$ &$\delta$ &$\mu_{\alpha}\cos(\delta)$&$\mu_{\delta}$&$parallax$&$g$ \\
\hline
&\multicolumn{2}{c|}{deg}&\multicolumn{2}{c|}{\masyr}&mas&mag\\
\hline
1&132.799&11.756&-10.86$\pm$0.11&-2.82$\pm$0.08&1.73$\pm$0.55&10.04\\
2&132.875&11.788&-10.94$\pm$0.13&-2.82$\pm$0.10&1.03$\pm$0.26&9.12\\

\hline
\hline
\end{tabular}\\
\end{table}

\begin{table}
\caption{The proper motions of M67 from the different catalogs}.\label{tab:pm_M67}
\centering
\begin{tabular}{c|c|c}
\hline
\hline
Catalog &$\mu_{\alpha}\cos(\delta)$&$\mu_{\delta}$\\
\hline
&\multicolumn{2}{c}{\masyr}\\
\hline
TGAS&-10.90$\pm$0.12&-2.82$\pm$0.09\\
GPS1&-10.54$\pm$0.14&-2.94$\pm$0.12\\
PPMXL&-7.20$\pm$0.18&-5.80$\pm$0.13\\
UCAC4&-9.00$\pm$0.27&-5.10$\pm$0.21\\

\hline
\hline
\end{tabular}\\
\end{table}

\subsubsection{Proper motion validation using distant Galactic stars}\label{sect:dstars}
We collect $\sim2200$ distant stars ($d>20$ kpc) from the literature \citep{xxx2008, xxx2014} with $13.5<m_r<17.5$, and calculate their proper motions. \HJT{These distant halo stars roughly have zero mean velocity in the galactocentric frame and large velocity dispersion, so they could be used for validation of GPS1 proper motions. But most stars} in the sample are located in the range of $20<d<40$~kpc, not \HJT{distant enough} for the Sun's reflex motion to be negligible. 
Therefore, we must correct for the Solar reflex motion before we use them for validation.

\begin{figure}[!t]
\centering
\includegraphics[width=0.45\textwidth, trim=0.0cm 0.0cm 0.0cm 0.0cm, clip]{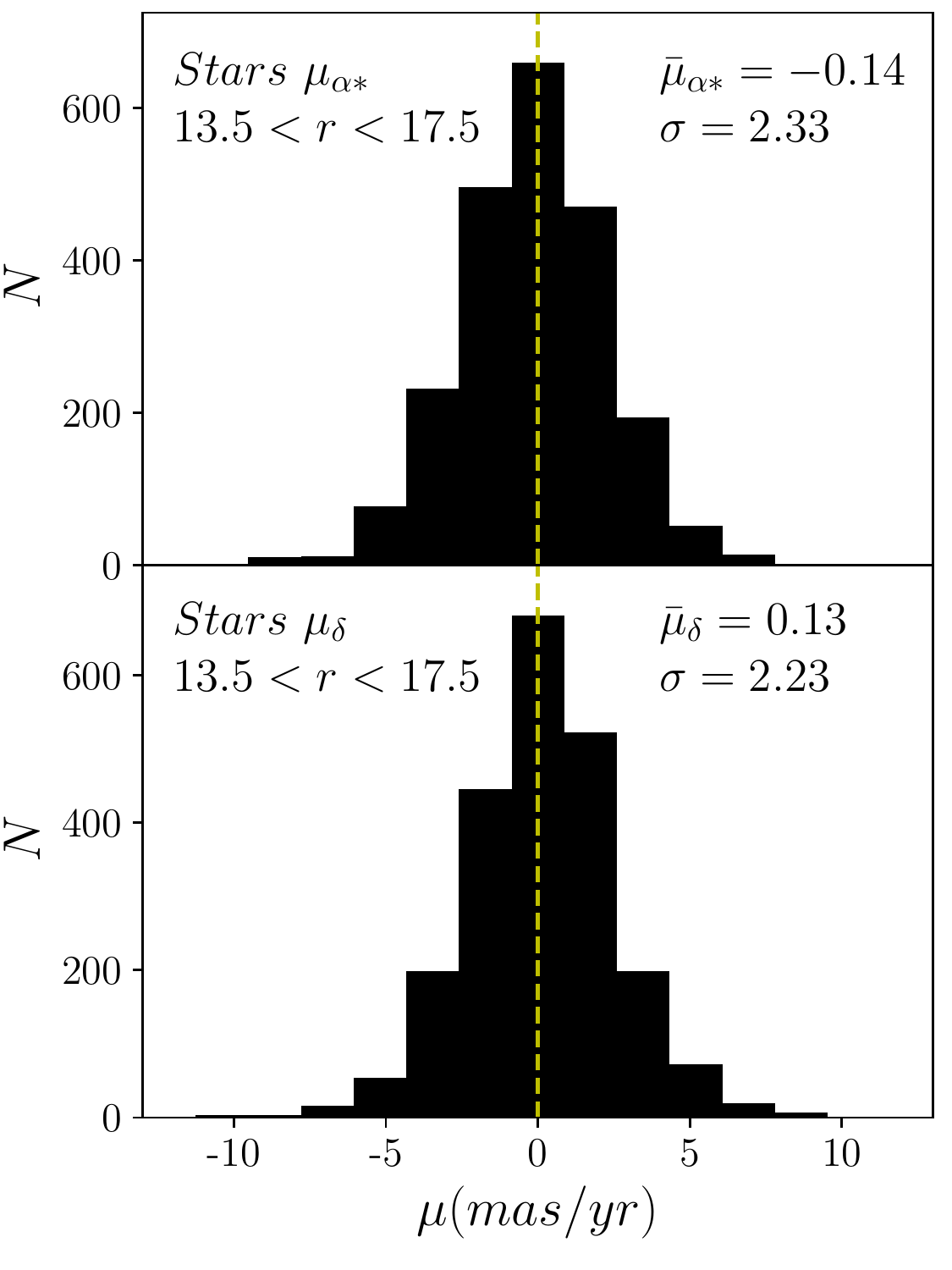}

\caption{Validation of GPS1 proper motions with distant halo stars. The solar motion
, $(U_\odot,V_\odot,W_\odot)=(9.58, 10.52, 7.01)$\,\kms\ \citep{tian2015}, 
has been removed, assuming an azimuthal LSR velocity of  $v_0=220$\,\kms. The light dashed lines denote zero \masyr ,
expected for a distant stellar halo of negligible rotation. 
}
   \label{fig:vali_dstars}
\end{figure}

Figure \ref{fig:vali_dstars} shows the histograms of the \mura\ (the top panel) and \mudec\ (the bottom panel), \HJT{where we have adopted} the solar motion as $(U_\odot,V_\odot,W_\odot)=(9.58, 10.52, 7.01)$\,\kms\ \citep{tian2015}, and the IAU recommended circular speed of LSR as $v_0=220$\,\kms, to remove the solar reflex motion.
The median values of the \mura\ and \mudec\ are -0.14 \masyr\ and 0.13 \masyr, and the dispersions are 2.33 \masyr\ and 2.23 \masyr, respectively. Accounting for this correction,  the mean
halo star proper motions are well within our accuracy estimate of 0.3 \masyr .

The velocity dispersion of the halo stars widens the distribution of proper motions. Using the distances provided by \citet{xxx2008, xxx2014}, one can calculate the median distance ($\sim$25\, kpc) of this sample. Supposing the velocity
dispersion in the halo is $\sim 100$ \kms\ \citep{deason2013}, then the corresponding proper motion is around 0.85\masyr, which indicates that the true {\it rms} of the proper motion \HJT{estimates} is $\sim 2$\masyr. This {\it rms} is slightly larger than 1.5\masyr\ as measured in Figure \ref{fig:sigmu_mr_GPS1} and \ref{fig:uncertanties_star}.

\subsection{Comparison with other proper motions}\label{sect:val_pal5}
\citet[][hereafter, FK15]{fk2015} obtained \HJT{high quality} proper motion measurement of Palomar 5, using SDSS and Large Binocular Camera (LBC) images. From the authors, we have obtained a proper motion sample of 1916 bright stars ($14.0<m_r<17.5$) \HJT{in the field of} Palomar 5. The proper motions \HJT{show a wide range} since most of the objects are field stars. 

\HJT{We then calculate the proper motions of stars in the same field, and cross-match the stars with the sample provided by FK15}. With the cross-matched 1887 stars, we compare our proper motions with different combinations, and also with PPMXL and UCAC4.

\begin{figure*}[!t]
\centering
\includegraphics[width=0.45\textwidth, trim=0.0cm 0.0cm 0.0cm 0.0cm, clip]{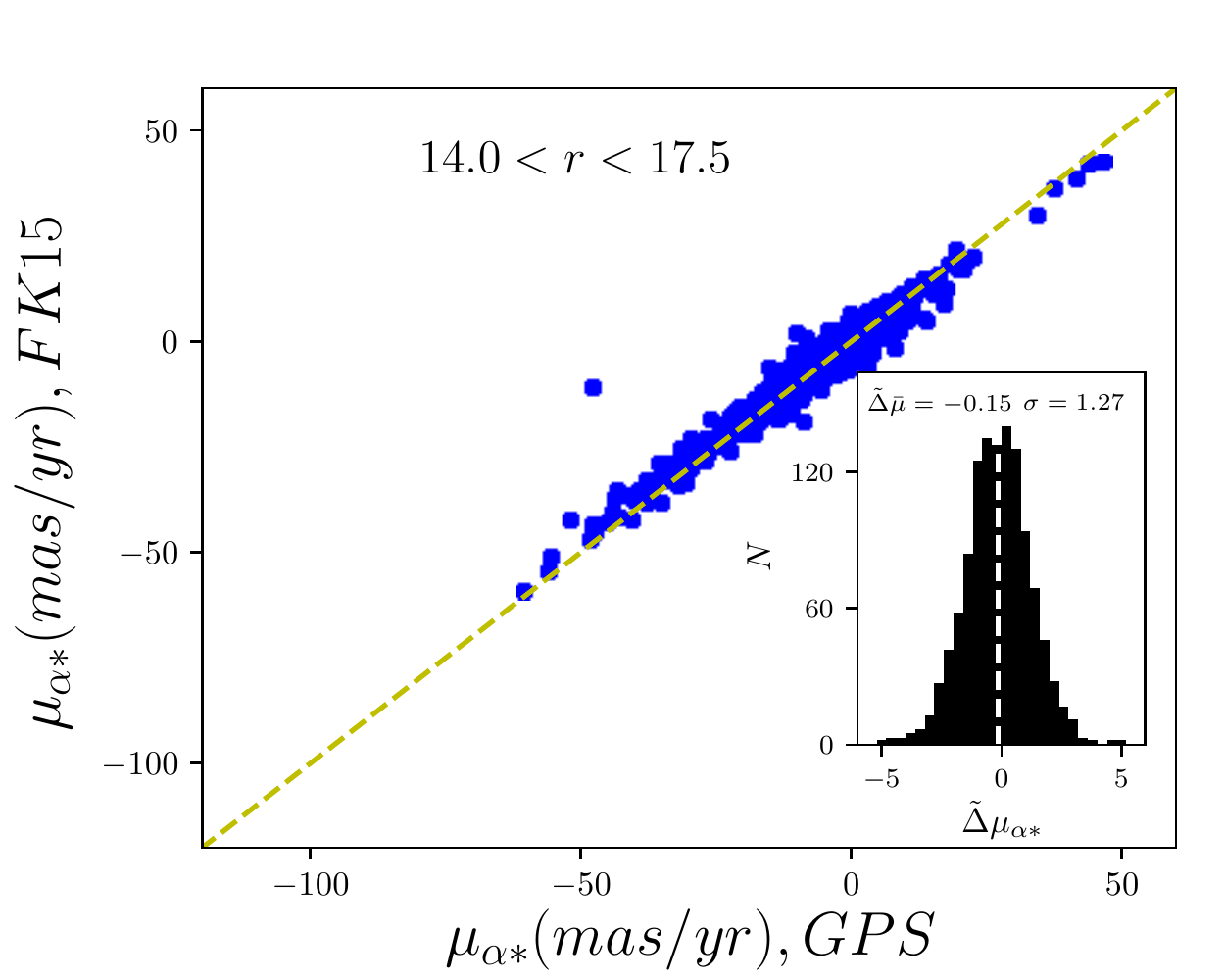}
\includegraphics[width=0.45\textwidth, trim=0.0cm 0.0cm 0.0cm 0.0cm, clip]{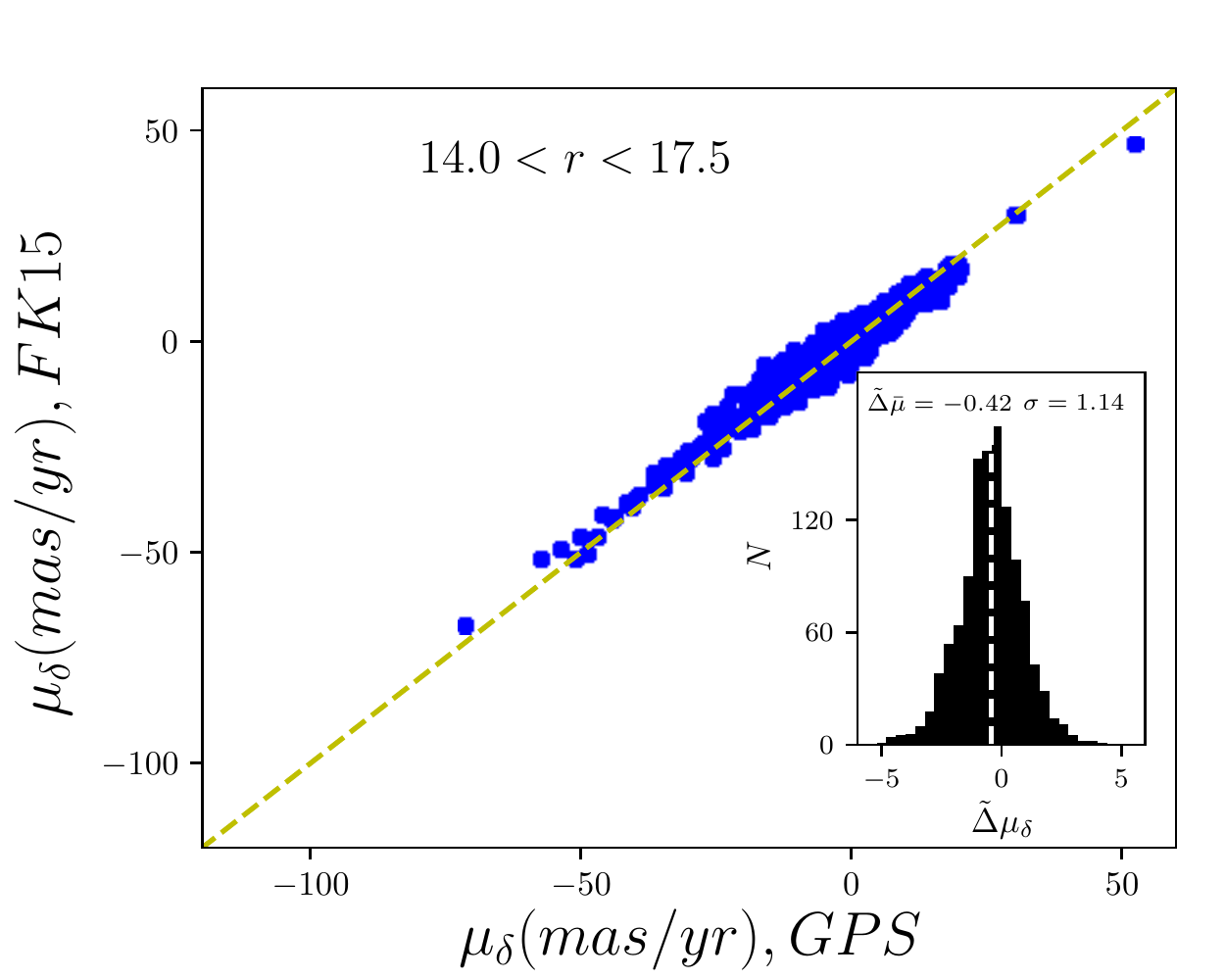}
\caption{Comparison of proper motions between GPS and FK15 for \mura\ (the left panel) and \mudec\ (the right panel), based on stars near the Palomar 5. The insets are histograms of the error-weighted difference between our proper motion and FK15. The median of the error-weighted differences (the white dashed line) for the \mura\ and \mudec\ are $-0.15\pm1.27$ and $-0.42\pm1.14$ (the absolute values: $-0.27\pm2.27$ \masyr and $-0.80\pm2.10$ \masyr), respectively. Note that the SDSS positions have been used in both proper motions estimates.
}
   \label{fig:vali_pal5}
\end{figure*}

Figure \ref{fig:vali_pal5} represents the comparison of proper motions between our GPS case and FK15 for \mura\ (the left panel) and \mudec\ (the right panel). The insets are the histograms of the error-weighted difference between the two, e,g. $\tilde{\Delta} \mu = (\mu_{ours} - \mu_{FK15})/\sqrt{\smash[b]{\epsilon_{\mu, ours}^2 + \epsilon_{\mu, FK15}^2}}$, where the two $\epsilon$ are the errors of our and FK15 proper motions. The median of the error-weighted differences (marked by the white dashed lines) for the \mura\ and \mudec\ are $-0.15\pm1.27$ and $-0.42\pm1.14$ (the absolute values: $-0.27\pm2.27$ \masyr and $-0.80\pm2.10$ \masyr), respectively. \HJT{The plot shows} that our proper motions are consistent with FK15 at the $1\sigma$ level. Note that their \mudec\ estimates are higher by $\sim0.8$ \masyr\ \HJT{than} ours, but the differences in \mura\ are not notable. We also compared the proper motions from the GP, PS1, PPMXL and UCAC4 with FK15, and found that the \mura\ are matched, just with much larger dispersions (3.0$\sim$3.5\masyr). But \mudec\ from FK15 is also higher by $\sim$1.0\masyr\ than that from GP, PS1, PPMXL, and UCAC4.

\section{\HJT{GPS1 Limitations}}

For the most part, GPS1 should constitute a catalog that is far more accurate and considerably more precise than existing catalogs of 
comparable size and depth, PPMXL and UCAC4. \HJT{In Section \ref{sec:val} we have described our catalog validation efforts}. Here, we discuss 
a number of regimes, where GPS1 has limitations beyond its \HJT{quoted} uncertainties, and where it should be used with caution.

\subsection{Proper Motions in Crowded regions}\label{sect:pm_crowed}

The reference frame calibration across the different surveys 
is difficult in crowded regions, for example near globular clusters, for two reasons: on the one hand, source crowding (the different surveys differ by a factor of $\ge 5$ in resolution) may lead to systematic errors in source centering. On the other hand, blended sources may be classified erroneously as extended sources, \HJT{which are then presumed to have zero proper motion}. If that happens too often in crowded regions, the mean proper motion of the stellar sample may inevitably be driven towards zero by our calibration approach.  

To explore these effects, we calibrated a sample of stars near the globular cluster M13, with two variants of bringing different epochs to the same reference frame. In one case we \HJT{use} objects classified as galaxies within the core radius. In the other, we do not. For these two cases, the final proper motions of stars close to the core region are significantly different. This indicates that misclassification of galaxies in a crowded core causes poor reference frame calibration. 
\HJT{If stars are too close, i.e., if their angular distance is smaller than 1.5 mas, the procedure that associates repeatedly observed positions with unique objects, will fail for one of the stars in such a pair. The proper motion of such a star will thus be incorrect.}

In order to reduce the impacts of crowding on positional calibration, we simply do not use the galaxies within the core radius of \HJT{known} globular clusters. We also use the bright galaxies in the Galactic plane to build a relatively reliable reference with PS1 for this study. In a future paper, Tian et. al. ({\it in preparation}) we plan a more complex calibration for stars in crowded regions with the aim of measuring proper motions for thousands of known star clusters and search for new star cluster candidates using the GPS1 catalog.


\subsection{The impact of DCR} \label{sect:dcr}
Refraction of the Earth's atmosphere varies with airmass and \HJT{wavelength, resulting in} differential chromatic refraction (DCR). Airmass depends on declination, right ascension, and the observational time and location of the observatory. \HJT{For PS1}, most observations are taken near meridian, so declination becomes a rough proxy for airmass. 

\begin{figure*}[!t]
\centering
\includegraphics[width=0.45\textwidth, trim=0.0cm 0.0cm 0.0cm 0.0cm, clip]{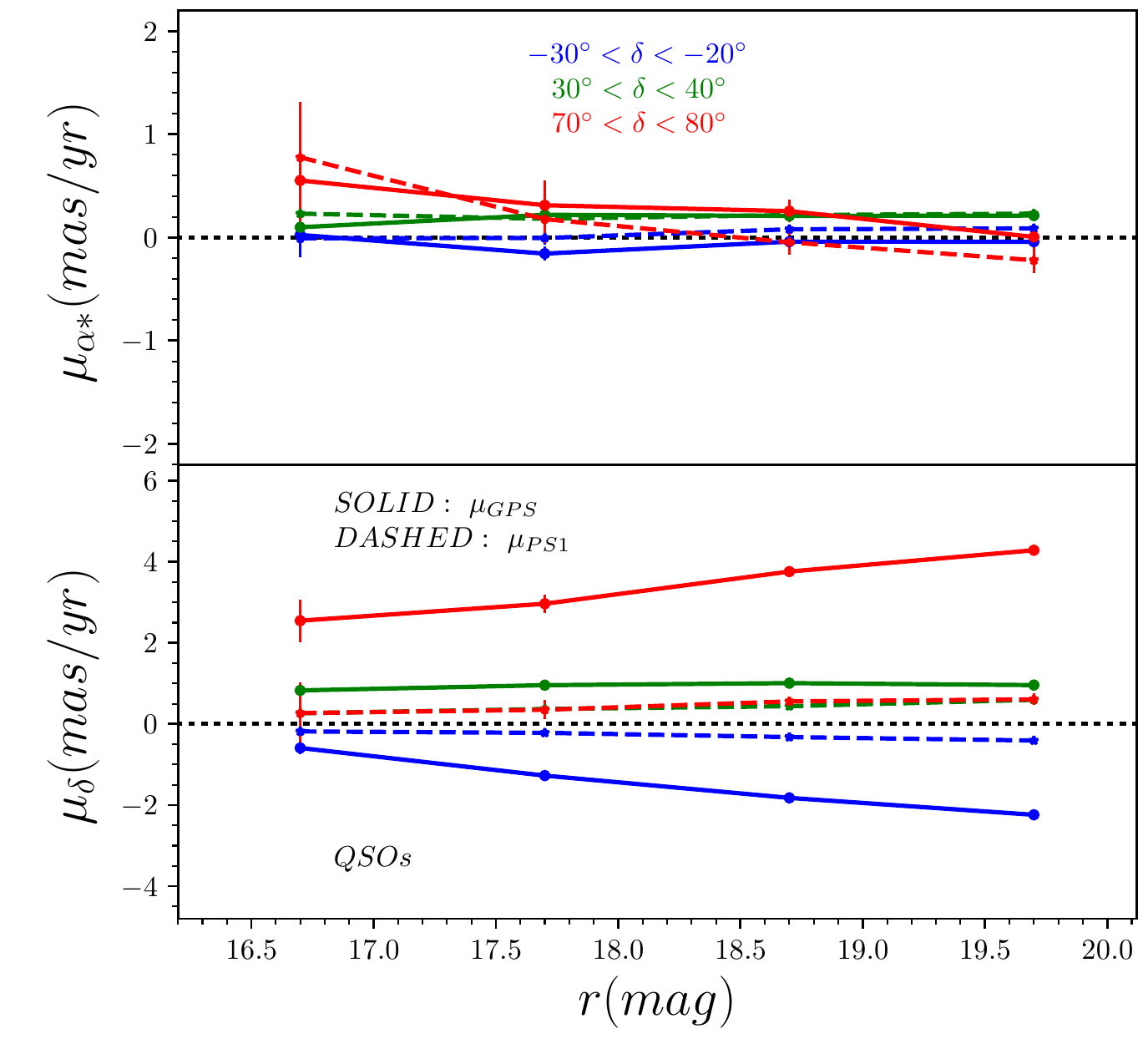}
\includegraphics[width=0.45\textwidth, trim=0.0cm 0.0cm 0.0cm 0.0cm, clip]{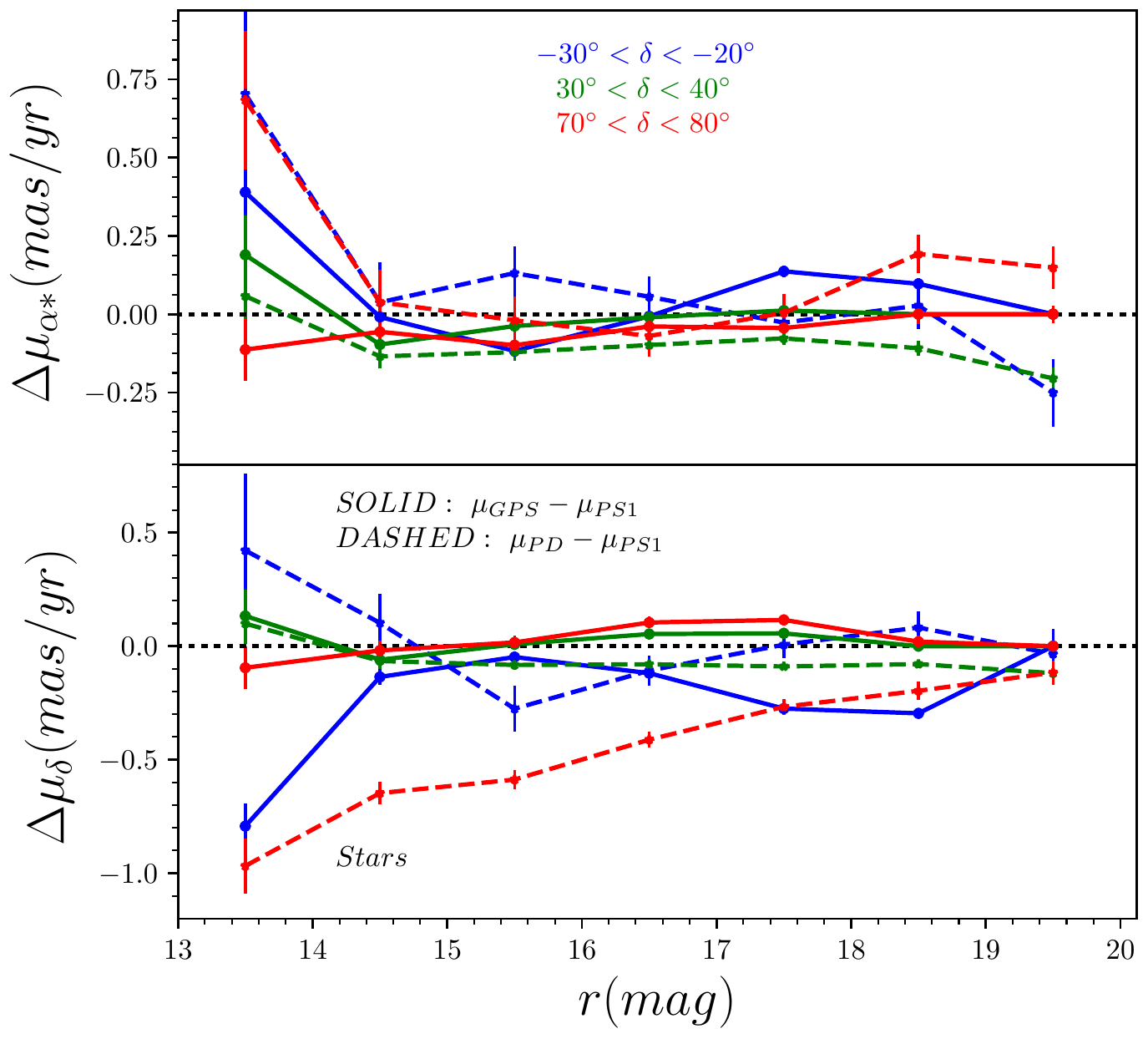}
\caption{The magnitude dependent and declination dependent DCR impacts on QSOs (the left panel) and stars (the right panel). The solid (dashed) curves represent the proper motions in GPS (PS1) in the left panel, and the difference of proper motions between GPS (PD) and PS1 modes in the right panel. The different colors in both the panels depict the cases of different declinations (blue for low, green for intermediate and red for high), while the black dotted lines mark the zero-level. 
}\label{fig:dcr}
\end{figure*}

The effect of DCR on QSOs is complex, due to their wide emission lines and a large range in redshifts  \citep{Kaczmarczik2009}. Moreover, Gaia, PS1 and SDSS surveys were conducted under different conditions: Gaia is a space-based telescope, and its observations are not affected by DCR; while PS1 and SDSS are ground-based telescopes and located in different places, so the two surveys suffer from DCR to a different extent. The combination of different surveys in the proper motion fit \HJT{may lead to complex DCR effects}. In order to figure out how the DCR affects the proper motions, we choose three samples of QSOs and stars in three declinations, and divide them into different magnitude bins. Figure \ref{fig:dcr} displays the magnitude and declination dependent impact of DCR \HJT{(reflected in non-zero proper motions)} on QSOs in the left panel and on stars in the right panel. 

For QSOs in the PS1 case, the observations near the meridian greatly reduce the impact of DCR on \mura\ (top-left, dashed lines). The deviation from zero (dot black line) is only $\sim -0.2 (+0.2)$\masyr\ in the low (high) declinations, particularly in the bins with $m_r>17.5$. However, the impact on \mudec\ is much more pronounced (bottom-left, dashed lines), the deviation is $\sim -0.5 $ $ (+0.5)$\masyr\ in the low (high) declination. The \HJT{tendency of} under (over) estimation of \mudec\ in the low (high) declination increases \HJT{as towards faint end}. \HJT{In the GPS case (solid lines), the impact of DCR on \mudec\ (bottom-left) turns to be worse}. The under (over) estimation can be up to $\sim$-4.0 (+4.0)\masyr\ in the low (high) declination, and the trends of the deviation from zero are almost same as the case in PS1. Gaia and SDSS detections, even if they were more "correct" than the PS1 measurements, apparently amplify the effect of DCR.

For stars we can still explore how the inclusion of Gaia and SDSS data affect the proper motion estimates, compared to PS1 only. We do this by analyzing the proper motion estimate differences, for example, $\mu_{GPS} - \mu_{PS1}$ (solid curves) and $\mu_{PD} - \mu_{PS1}$ (dashed curves), shown in the right panel of Figure \ref{fig:dcr}. The difference between GPS and PS1 is very small ($\Delta\mu_{\alpha}<0.2$\masyr, $\Delta\mu_{\delta}<0.5$\masyr) for both \mura\ (top-right) and \mudec\ (bottom-right) except in the bin of bright stars where the star number is small. It indicates that Gaia and SDSS do not amplify the DCR impacts on stars \THJ{unlike} QSOs. \HJT{The dashed lines demonstrate that the SDSS detections change the inferred proper motions of PS1 significantly at high declinations. The shifts on \mudec\ of PS1 caused by SDSS detections (bottom-right, red dashed curve) are linearly dependent on magnitude, and $\Delta\mu_{\delta}$ can deviate as much as $\sim-0.9$\masyr}. Interestingly, Gaia seems to \HJT{replicate} this kind of significant shifts caused by SDSS, as shown by the red solid curve in the bottom right panel.

\begin{figure}[!t]
\centering
\includegraphics[width=0.45\textwidth, trim=0.0cm 0.0cm 0.0cm 0.0cm, clip]{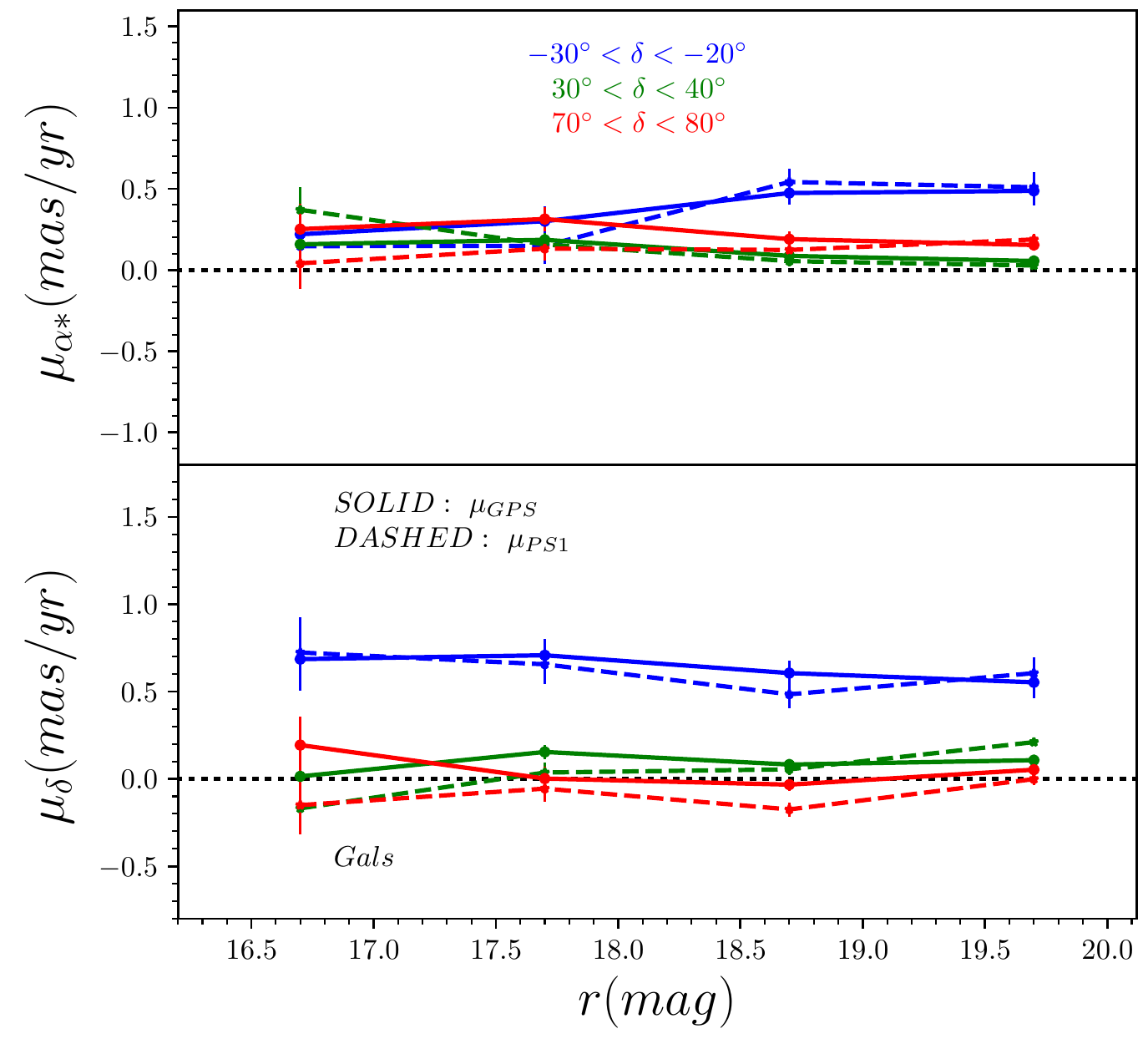}
\caption{Same as Figure \ref{fig:dcr}, but depicts the case of proper motions of galaxies in different magnitudes and declination bins. The median proper motions are $\sim 0$ \masyr\ ($<0.2$\masyr), except in the low declination, where they are non-zero ($<0.5$\masyr). 
}
 \label{fig:dcr_gals}
\end{figure}

We also investigate the case of galaxies, but do not detect any significant impacts due to DCR. The median proper motions of galaxies are around 0.0 \masyr\ ($<0.2$\masyr), except at low declinations where they are non-zero ($<0.5$\masyr), as displayed in Figure \ref{fig:dcr_gals}.

In general, the DCR impacts the proper motions of QSO in a complex and noticeable manner. Therefore, QSOs are not ideal sources for the validation of proper motions, as discussed in Section \ref{sect:val_qso}. However, DCR has a more benign influence on stars and galaxies.




\subsection{The origin of the magnitude and declination dependent offsets in predicted Gaia positions}\label{sect:moffset}
In Section \ref{sec:mag_offset}, we reported the positional offsets between the predicted, from PS1, and the originally observed Gaia positions. As shown in Figure \ref{fig:offset2}, the positional offsets of stars are significantly \THJ{dependent on magnitude and declination}. Positions can have \THJ{offsets} of up to $\sim\pm10$ mas. 

\citet{Magnier2016} point out that the position of PS1 sources depends on flux and may be affected by imperfect DCR corrections. Due to charge leakage, bright stars are offset on PS1 camera CCDs relative to faint stars. This leakage is \HJT{stronger} in the case of brighter stars. This effect was first identified by \citet{Koppenhofer2011}. The DCR is a typical declination dependent effect that arises due to variation of airmass in the direction of declination for observations near meridian. The combined impact of the two effects might induce spurious proper motions from PS1, with which the predicted Gaia positions could be magnitude and declination dependent. Although these two systematic effects were corrected in \citet{Magnier2016}, that correction may be imperfect.

\section{Conclusions}\label{sect:conclusions}
The PS1, Gaia, SDSS, and 2MASS surveys have collected positions for billions of stars, with precise relative astrometry across a baseline of $\sim 15$ years. By combining them, we build a catalog of proper motions for $\sim$ 350 million point sources brighter than $m_r\sim20$\,mag, across three quarters of the sky. The systematic error (i.e. accuracy) is $<0.3$ \masyr\, and the typical uncertainty in the proper motion of a single source is $\sim 1.5$\masyr (for sources brighter than $m_r=18$).

Our analysis required that the cataloged source positions of all surveys at all separate epochs be brought to a common reference frame.
We accomplished this by requiring that galaxies have zero proper motion and that angular motions of stars on the sky are essentially linear. 
We verified that this approach leads to proper motion estimates of the precision and accuracy stated above. There are several important exceptions that we discuss, in particular QSOs and crowded fields. 

We compare GPS1 with published large scale proper motion catalogs in Section \ref{sect:val_pal5}: the accuracy of GPS1 ($<0.3$ \masyr) is $\sim$ 10 times better than PPMXL and UCAC4 ($>2.0$ \masyr), and the precision ($\sim 1.5$ \masyr) is $\sim$ 4 times better than PPMXL and UCAC4 ($\sim 6.0$ \masyr).

Until Gaia DR2, GPS1 should provide a valuable resource for kinematic studies of the Milky Way.

\acknowledgements
The authors thank Bin Yue, Andy Gould for the helpful discussions, \HJT{and Tobias Fritz for providing the proper motions of stars in the field of Palomar 5}. H.-J.T. acknowledges the National Natural Science Foundation of China (NSFC) under grants 11503012, U1331202 and the fellowships from China Scholarship Council (CSC) and Max Planck Institute for Astronomy. B.S. and H.-W.R. acknowledge funding from the European Research Council under the European Unions Seventh Framework Programme (FP 7) ERC Grant Agreement n. [321035]. B.G. acknowledge funding from Sonderforschungsbereich SFB 881 "The Milky Way System" (subproject A3 and B6) of the German Research Foundation (DFG). The Pan-STARRS1 Survey (PS1) has been made possible through contributions of the Institute for Astronomy at the University of Hawaii, Pan-STARRS Project Office, Max-Planck Society and its participating institutes, specifically  Max Planck Institute for Astronomy, Heidelberg and Max Planck Institute for Extraterrestrial Physics, Garching, Johns Hopkins University, Durham University, University of Edinburgh, Queen's University Belfast, Harvard-Smithsonian Center for Astrophysics, Las Cumbres Observatory Global Telescope Network Incorporated, National Central University of Taiwan, Space Telescope Science Institute, National Aeronautics and Space Administration under Grant No. NNX08AR22G issued through the Planetary Science Division of the NASA Science Mission Directorate, the National Science Foundation under Grant No. AST-1238877, University of Maryland, Eotvos Lorand University and Los Alamos National Laboratory. This work has made use of data from the European Space Agency (ESA)
mission {\it Gaia} (\url{https://www.cosmos.esa.int/gaia}), processed by
the {\it Gaia} Data Processing and Analysis Consortium (DPAC,
\url{https://www.cosmos.esa.int/web/gaia/dpac/consortium}). Funding
for the DPAC has been provided by national institutions, in particular
the institutions participating in the {\it Gaia} Multilateral Agreement.

\bibliographystyle{apj}

\bibliographystyle{yahapj}

\end{document}